\documentstyle[prd,epsf,aps,eqsecnum]{revtex}
\begin{document}
\twocolumn[%
\hsize\textwidth\columnwidth\hsize\csname@twocolumnfalse\endcsname
\title{%
\hfill{\normalsize\vbox{\hbox{September 2000} \hbox{DPNU-00-29}  }}\\
\bf Wilsonian Matching of Effective Field Theory \\
with Underlying QCD}
\author{{\bf Masayasu Harada} and {\bf Koichi Yamawaki}}
\address{Department of Physics, Nagoya University
Nagoya, 464-8602, Japan.}
\maketitle

\begin{abstract}
We propose a novel way of matching effective field theory with the
underlying QCD in the sense of a Wilsonian renormalization group
equation 
(RGE).  We derive Wilsonian matching conditions between current
correlators obtained by the operator product expansion in QCD and
those by the hidden local symmetry (HLS) model.  This
determines without much ambiguity the bare parameters of the HLS at
the cutoff scale in terms of the QCD parameters.  Physical quantities
for the $\pi$ and $\rho$ system are calculated by the Wilsonian RGE's
from the bare parameters in remarkable agreement with the experiment.
\end{abstract}
\vskip1pc]

\section{Introduction}

Recently the concept of the Wilsonian renormalization group equation
(RGE) has 
become fashionable in the context of 
matching effective field theories (EFT's)
with underlying gauge theories  
to study the phase structure of supersymmetric (SUSY) gauge 
theories~\cite{Seiberg}.
However, no attempt has been made to match the EFT
with the underlying (non-SUSY) QCD in the sense of a Wilsonian RGE
which now includes 
{\it quadratic divergences} in addition to the logarithmic ones in the
RGE flow of the EFT. It would be reasonable to
consider the effective theory under an ordinary RGE 
with just a logarithmic divergence in the situation
where spontaneous chiral symmetry breaking is always 
granted from the beginning as in
QCD with the number of almost massless flavors being $N_f =3$. 
Actually, the logarithmic RGE is blind about the change of phase.

In a previous paper\cite{HY:letter} we actually
demonstrated that the inclusion
of a quadratic divergence in the Wilsonian sense in the EFT
does give rise to 
chiral symmetry restoration by its own dynamics for large $N_f$ under  
certain conditions,
based on the Hidden Local Symmetry (HLS) Lagrangian~\cite{BKUYY,BKY} 
which successfully incorporates $\rho$ and its 
flavor partners in the chiral Lagrangian.
Chiral symmetry restoration for large $N_f$ QCD is a notable
phenomenon observed by various methods such as lattice
simulations~\cite{lattice},
the Schwinger-Dyson equation approach~\cite{SD}, 
the dispersion relation~\cite{OZ},
instanton calculations~\cite{VS}, etc.

In this paper, we shall propose a novel way of {\it matching the EFT
with the underlying QCD with $N_f=3$}
in the sense of a Wilsonian RGE,
namely, including {\it quadratic divergences} in the EFT (``Wilsonian
matching'').  By this we demonstrate that inclusion of the quadratic
divergence is important even for phenomenology in the $N_f=3$
QCD.
The basic tool of Wilsonian matching is the {\it Operator Product
Expansion} (OPE) of QCD for the axialvector and vector current
correlators, which
are equated with those from the EFT at the matching scale $\Lambda$.
{\it This determines without much ambiguity
the bare parameters of the EFT defined at the scale $\Lambda$ 
in terms of the QCD parameters.}
Physical quantities for the $\pi$ and $\rho$ system
are calculated by the Wilsonian RGE's
from the bare parameters
in remarkable agreement with experiment.

\section{Hidden Local Symmetry}

Let us first describe the EFT, the HLS model based on the
$G_{\rm global} \times H_{\rm local}$ symmetry, where
$G = \mbox{SU($N_f$)}_{\rm L} \times 
\mbox{SU($N_f$)}_{\rm R}$  is the 
global chiral symmetry and 
$H = \mbox{SU($N_f$)}_{\rm V}$ is the HLS.
(The flavor symmetry is given by the diagonal sum of $G_{\rm global}$
and $H_{\rm local}$.)
The basic quantities are the gauge boson $\rho_\mu$ 
of the HLS and two 
SU($N_f$)-matrix-valued variables $\xi_{\rm L}$ and 
$\xi_{\rm R}$. They transform as
\begin{equation}
\xi_{\rm L,R}(x) \rightarrow \xi_{\rm L,R}^{\prime}(x) =
h(x) \xi_{\rm L,R}(x) g^{\dag}_{\rm L,R}
\ ,
\end{equation}
where $h(x) \in H_{\rm local}$ and 
$g_{\rm L,R} \in G_{\rm global}$.
These variables are parametrized as
\begin{equation}
\xi_{\rm L,R} = e^{i\sigma/F_\sigma} e^{\mp i\pi/F_\pi}
\ ,
\label{xiLR}
\end{equation}
where $\pi = \pi^a T_a$
denotes the Nambu-Goldstone (NG) bosons associated with 
the spontaneous breaking of $G$ chiral symmetry and 
$\sigma = \sigma^a T_a$
the NG bosons absorbed into the gauge bosons.
$F_\pi$ and $F_\sigma$ are relevant decay constants, and
the parameter $a$ is defined as
\begin{equation}
a \equiv F_\sigma^2/F_\pi^2
\ .
\end{equation}
Here $\pi$ denotes the pseudoscalar NG bosons associated with the
chiral $\mbox{SU($N_f$)}_L \times \mbox{SU($N_f$)}_R$ symmetry
and $\rho$ the HLS gauge bosons even though we fix $N_f = 3$.
The covariant derivatives of $\xi_{\rm L,R}$ are defined by
\begin{equation}
D_\mu \xi_{\rm L} =
\partial_\mu \xi_{\rm L} - i g \rho_\mu \xi_{\rm L}
+ i \xi_{\rm L} {\cal L}_\mu
\ ,
\label{covder}
\end{equation}
and similarly with the replacement 
${\rm L} \leftrightarrow {\rm R}$,
${\cal L}_\mu \leftrightarrow {\cal R}_\mu$,
where $g$ is the HLS gauge coupling.
${\cal L}_\mu$ and ${\cal R}_\mu$ denote the external gauge fields
gauging the $G_{\rm global}$ symmetry.

The HLS Lagrangian is given by~\cite{BKUYY,BKY}
\begin{equation}
{\cal L} = F_\pi^2 \, \mbox{tr} 
\left[ \hat{\alpha}_{\perp\mu} \hat{\alpha}_{\perp}^\mu \right]
+ F_\sigma^2 \, \mbox{tr}
\left[ 
  \hat{\alpha}_{\parallel\mu} \hat{\alpha}_{\parallel}^\mu
\right]
+ {\cal L}_{\rm kin}(\rho_\mu) \ ,
\label{Lagrangian}
\end{equation}
where ${\cal L}_{\rm kin}(\rho_\mu)$ denotes the kinetic term of
$\rho_\mu$ 
and
\begin{eqnarray}
&&
\hat{\alpha}_{\stackrel{\perp}{\scriptscriptstyle\parallel}}^\mu =
\left( 
  D_\mu \xi_{\rm L} \cdot \xi_{\rm L}^\dag \mp
  D_\mu \xi_{\rm R} \cdot \xi_{\rm R}^\dag 
\right)
/ (2i) \ .
\end{eqnarray}

\section{Renormalization Group Equations in the Wilsonian Sense}

In Ref.~\cite{HY:letter}
the quadratic divergence was identified with the presence of
poles of ultraviolet origin at $n=2$ in the dimensional
regularization~\cite{Veltman}.
The resultant RGE's for $F_\pi^2$, $a$ and $g^2$ are given 
by~\cite{HY:letter}
\begin{eqnarray}
&&
\mu \frac{d F_\pi^2}{d\mu} =
C \left[ 3 a^2 g^2 F_\pi^2 + 2 (2-a) \mu^2 \right] \ ,
  \label{RGE for Fpi2}
\nonumber\\
&&
\mu \frac{d a}{d\mu} = - 
C
(a-1)
\left[ 3 a (a+1) g^2 - (3a-1) \frac{\mu^2}{F_\pi^2} \right] \ ,
  \label{RGE for a}
\nonumber\\
&&
\mu \frac{d g^2}{d\mu} = - 
C
\frac{87 - a^2}{6} g^4 \ ,
  \label{RGE for g2}
\end{eqnarray}
where $C = N_f/[2(4\pi)^2]$ and
$\mu$ is the renormalization scale.
We note here that the above RGE's agree with those obtained in
Ref.~\cite{HY} when we neglect quadratic divergences.
A detailed derivation of the above RGE's is given in 
Appendixes~\ref{sec:BGFM} and \ref{sec:RGE}.

In addition to the leading-order terms (\ref{Lagrangian})
we need to include the 
${\cal O}(p^4)$ higher derivative terms in the present 
analysis (see Appendix~\ref{sec:DEH}).
The relevant terms are given by~\cite{Tanabashi}
\begin{equation}
z_1 \mbox{tr}
\left[ \hat{\cal V}_{\mu\nu} \hat{\cal V}^{\mu\nu} \right] + 
z_2 \mbox{tr}
\left[ \hat{\cal A}_{\mu\nu} \hat{\cal A}^{\mu\nu} \right] 
+ 
g z_3 \mbox{tr}
\left[ \hat{\cal V}_{\mu\nu} \rho^{\mu\nu} \right]
\ ,
\label{z term}
\end{equation}
where
\begin{eqnarray}
\hat{\cal A}_{\mu\nu} &=&
\left( \xi_{\rm R} {\cal R}_{\mu\nu} \xi_{\rm R}^\dag
- \xi_{\rm L} {\cal L}_{\mu\nu} \xi_{\rm L}^\dag
\right)/2
\ ,\nonumber\\
\hat{\cal V}_{\mu\nu} &=&
\left( \xi_{\rm R} {\cal R}_{\mu\nu} \xi_{\rm R}^\dag
+ \xi_{\rm L} {\cal L}_{\mu\nu} \xi_{\rm L}^\dag
\right)/2
\ ,
\label{def:AV}
\end{eqnarray}
with ${\cal R}_{\mu\nu}$ and ${\cal L}_{\mu\nu}$ being the field
strengths of ${\cal R}_{\mu}$ and ${\cal L}_{\mu}$.
Here $\rho_{\mu\nu}$
is the gauge field strength of the HLS gauge boson.
Since there are no quadratically divergent corrections to the 
parameters $z_1$, $z_2$ and $z_3$,
we calculate the RGE's from the logarithmic divergences
listed in Ref.~\cite{Tanabashi}:
\begin{eqnarray}
&&
\mu \frac{d z_1}{d \mu} =
\frac{N_f}{(4\pi)^2} \frac{5-4a+a^2}{24} \ , 
\quad
\mu \frac{d z_2}{d \mu} =
\frac{N_f}{(4\pi)^2} \frac{a}{12} \ ,
\nonumber\\
&& \qquad
\mu \frac{d z_3}{d \mu} =
\frac{N_f}{(4\pi)^2} \frac{1+2a-a^2}{12} \ .
\label{RGE z}
\end{eqnarray}

\section{Wilsonian Matching}

Now we propose a Wilsonian matching of the EFT with the underlying
QCD:
We determine the bare parameters as boundary values of the Wilsonian
RGE's (\ref{RGE for g2}) and (\ref{RGE z}) including quadratic
divergences
by matching the HLS with the OPE in QCD at the
matching scale $\Lambda$.

Let us look at axialvector and vector current correlators.  They are
well described by the tree contributions with including
${\cal O}(p^4)$ terms 
when the momentum is around the matching scale, $Q^2 \sim \Lambda^2$.
The resultant expressions of the correlators
are given by
\begin{eqnarray}
\Pi_A^{\rm(HLS)}(Q^2) &=&
\frac{F_\pi^2(\Lambda)}{Q^2} - 2 z_2(\Lambda)
\ ,
\nonumber\\
\Pi_V^{\rm(HLS)}(Q^2) &=&
\frac{
  F_\sigma^2(\Lambda) \left[ 1 - 2 g^2(\Lambda) z_3(\Lambda) \right]
}{
  M_v^2(\Lambda) + Q^2
} 
- 2 z_1(\Lambda)
\ ,
\nonumber\\
&&
\label{Pi A V HLS}
\end{eqnarray}
where we defined
\begin{equation}
M_v^2(\Lambda) \equiv g^2(\Lambda) F_\sigma^2(\Lambda)
\ .
\end{equation}
The same correlators are evaluated by the OPE 
up until ${\cal O}(1/Q^6)$~\cite{SVZ}:
\begin{eqnarray}
&&
\Pi_A^{\rm(QCD)}(Q^2) = \frac{1}{8\pi^2}
\Biggl[
  - \left( 1 + \frac{\alpha_s}{\pi} \right) \ln \frac{Q^2}{\mu^2}
\nonumber\\
&& \quad
  + \frac{\pi^2}{3} 
    \frac{
      \left\langle 
        \frac{\alpha_s}{\pi} G_{\mu\nu} G^{\mu\nu}
      \right\rangle
    }{ Q^4 }
  + \frac{\pi^3}{3} \frac{1408}{27}
    \frac{\alpha_s \left\langle \bar{q} q \right\rangle^2}{Q^6}
\Biggr]
\ ,
\nonumber\\
&&
\Pi_V^{\rm(QCD)}(Q^2) = \frac{1}{8\pi^2}
\Biggl[
  - \left( 1 + \frac{\alpha_s}{\pi} \right) \ln \frac{Q^2}{\mu^2}
\nonumber\\
&& \quad
  + \frac{\pi^2}{3} 
    \frac{
      \left\langle 
        \frac{\alpha_s}{\pi} G_{\mu\nu} G^{\mu\nu}
      \right\rangle
    }{ Q^4 }
  - \frac{\pi^3}{3} \frac{896}{27}
    \frac{\alpha_s \left\langle \bar{q} q \right\rangle^2}{Q^6}
\Biggr]
\ ,
\label{Pi A V OPE}
\end{eqnarray}
where $\mu$ is the renormalization scale of QCD.

We require that current correlators in the HLS
in Eq.~(\ref{Pi A V HLS})
can be matched with those in QCD in Eq.~(\ref{Pi A V OPE}).
Note that both $\Pi_A^{\rm(QCD)}$ and $\Pi_V^{\rm(QCD)}$ explicitly
depend on $\mu$~\cite{foot:mudep}.
However, the difference between two correlators has no
explicit dependence on $\mu$~\cite{foot:OPE}.
Thus our first Wilsonian matching condition is given by
\begin{eqnarray}
&&
\frac{F_\pi^2(\Lambda)}{\Lambda^2} -
\frac{F_\sigma^2(\Lambda)\left[ 1 - 2 g^2(\Lambda)z_3(\Lambda)\right]
}{\Lambda^2 + M_v^2(\Lambda)}
- 2 \left[ z_2(\Lambda) - z_1(\Lambda) \right]
\nonumber\\
&& \quad
=
\frac{32\pi}{9}
\frac{\alpha_s \left\langle \bar{q} q \right\rangle^2}{\Lambda^6}
\ .
\label{match z}
\end{eqnarray}

We also require that the first derivative of $\Pi_A^{\rm(HLS)}$ in 
Eq.~(\ref{Pi A V HLS}) match that of $\Pi_A^{\rm(QCD)}$ in 
Eq.~(\ref{Pi A V OPE}), and similarly for $\Pi_V$'s.
This requirement gives two Wilsonian matching
conditions
\begin{eqnarray}
&&
\frac{F_\pi^2(\Lambda)}{\Lambda^2} 
= \frac{1}{8\pi^2}
\Biggl[
  1 + \frac{\alpha_s}{\pi}
\nonumber\\
&& \quad
  + \frac{2\pi^2}{3} 
    \frac{
      \left\langle 
        \frac{\alpha_s}{\pi} G_{\mu\nu} G^{\mu\nu}
      \right\rangle
    }{ \Lambda^4 }
  + \pi^3\, \frac{1408}{27}
    \frac{\alpha_s \left\langle \bar{q} q \right\rangle^2}%
    {\Lambda^6}
\Biggr]
\ ,
\label{match A}
\\
&&
\frac{F_\sigma^2(\Lambda)}{\Lambda^2} 
\frac{\Lambda^4 \left[ 1 - 2 g^2(\Lambda) z_3(\Lambda) \right]
}{\left( \Lambda^2 + M_v^2(\Lambda) \right)^2}
= \frac{1}{8\pi^2}
\Biggl[
  1 + \frac{\alpha_s}{\pi}
\nonumber\\
&& \quad
  + \frac{2\pi^2}{3} 
    \frac{
      \left\langle 
        \frac{\alpha_s}{\pi} G_{\mu\nu} G^{\mu\nu}
      \right\rangle
    }{ \Lambda^4 }
  - \pi^3\, \frac{896}{27}
    \frac{\alpha_s \left\langle \bar{q} q \right\rangle^2}%
    {\Lambda^6}
\Biggr]
\ .
\label{match V}
\end{eqnarray}

The above three equations (\ref{match z})--(\ref{match V}) 
are the Wilsonian matching conditions, which we
propose in this paper.

The right-hand sides in Eqs.~(\ref{match z})--(\ref{match V})
are directly determined from QCD.
First note that the matching scale $\Lambda$ must be smaller than the
mass of the $a_1$ meson which is not included in our effective theory,
whereas $\Lambda$ has to be big enough for the OPE to be valid.
Here we use 
\begin{equation}
\Lambda = 1.1\,,\  1.2\,\mbox{GeV} \ .
\label{matching scale}
\end{equation}
To determine the current correlators from the OPE we use
\begin{eqnarray}
&&
\left\langle \frac{\alpha_s}{\pi} G_{\mu\nu} G^{\mu\nu}
\right\rangle = 0.012 \,\mbox{GeV}^4 \ , 
\nonumber\\
&&\left\langle \bar{q} q \right\rangle_{\rm 1\,GeV} = - 
\left(\mbox{0.25\,GeV}\right)^3
\ ,
\end{eqnarray}
shown in Ref.~\cite{SVZ}
and 
\begin{equation}
\Lambda_{\rm QCD} = 350\ , 400\,\mbox{MeV}
\end{equation}
as typical values.
We use one-loop running to
estimate $\alpha_s(\Lambda)$ and 
$\left\langle \bar{q} q \right\rangle_\Lambda$.

\section{Determination of the Bare Parametes of the HLS Lagrangian}
\label{sec:DBPHL}

Then the bare parameters $F_\pi(\Lambda)$, $a(\Lambda)$,
$g(\Lambda)$, $z_3(\Lambda)$ and $z_2(\Lambda)-z_1(\Lambda)$
can be determined
through the Wilsonian matching conditions.
Actually, the Wilsonian matching
conditions in Eqs.~(\ref{match z})--(\ref{match V}) 
are not enough to determine all the relevant bare
parameters.  We therefore use the on-shell pion decay constant
$F_\pi(0)=88$\,MeV in the chiral limit~\cite{GL} and the
$\rho$ mass $m_\rho = 770$\,MeV as inputs.
The mass of $\rho$
is determined by the on-shell condition
\begin{equation}
m_\rho^2 = a(m_\rho) g^2(m_\rho) F_\pi^2(m_\rho)
\ .
\end{equation}
Below the $m_\rho$ scale, 
$\rho$ decouples 
and hence $F_\pi^2$ runs by the $\pi$-loop effect 
alone.~\cite{foot:piloop}
Since the parameter $F_\pi(\mu<m_\rho)$ does not smoothly
connect to $F_\pi(\mu>m_\rho)$ at the $m_\rho$ scale,
we need to include a finite renormalization effect 
(see Appendix~\ref{sec:RGE})
\begin{equation}
\left[ F_\pi^{(\pi)}(m_\rho) \right]^2 =
F_\pi^2(m_\rho) + 
\frac{N_f}{(4\pi)^2} \frac{a(m_\rho)}{2} m_\rho^2 \ ,
\label{finite renormalization}
\end{equation}
where $F_\pi^{(\pi)}(\mu)$ runs by the loop effect of $\pi$ for
$\mu<m_\rho$.

The resultant values of all the bare parameters of the HLS are shown
in Table~I together with those at $\mu=m_\rho$.
\begin{center}
\vspace{0.2cm}
\small
\begin{tabular}{|c||c|c|c|c|c|}
\hline
$\mu$ & $F_\pi(\mu)$ & $a(\mu)$ & $g(\mu)$ & $z_3(\mu)$ 
  & $z_2(\mu)-z_1(\mu)$ \\
\hline
$\Lambda$ & 0.149 & 1.19 & 3.69 & -5.23$\times10^{-3}$ 
  & -1.03$\times10^{-3}$ \\
$m_\rho$  & 0.110 & 1.22 & 6.33 & -6.34$\times10^{-3}$ 
  & -1.24$\times10^{-3}$ \\
\hline
\end{tabular}
\\
\vspace{0.2cm}
\begin{minipage}{8.3cm}
TABLE~I. \ 
Five parameters of the HLS at $\mu=\Lambda$ and $m_\rho$ for 
$\Lambda_{\rm QCD} = 400$\,MeV and
$\Lambda=1.1$\,GeV.
The unit of $F_\pi$ is GeV.
\end{minipage}
\vspace{0.2cm}
\end{center}

\section{Predictions}

Now that we have completely specified the bare Lagrangian, we can
predict the following physical quantities by the Wilsonian RGE's
including the quadratic divergences, Eqs.~(\ref{RGE for g2}) and
(\ref{RGE z}).

The $\rho$-$\gamma$ mixing strength:

The second term in Eq.~(\ref{Lagrangian}) gives the mass mixing
between $\rho$ and the external field of $\gamma$.
The third term in Eq.~(\ref{z term}) gives the kinetic mixing.
Combining these two at the on-shell of $\rho$ leads to the
$\rho$-$\gamma$ mixing strength:
\begin{equation}
g_\rho = g(m_\rho) F_\sigma^2(m_\rho) 
\left[ 1 - g^2(m_\rho) z_3(m_\rho) \right] \ .
\label{g rho}
\end{equation}

The Gasser-Leutwyler's parameter $L_{10}$~\cite{GL}:

The relation between $L_{10}$ and the parameters of the HLS at 
$m_\rho$ scale is given by~\cite{Tanabashi}
\begin{eqnarray}
&&
L_{10}(m_\rho) =
- \frac{1}{4g^2(m_\rho)} 
\nonumber\\
&& \quad
+ \frac{z_3(m_\rho) - z_2(m_\rho) + z_1(m_\rho)}{2}
+ \frac{N_f}{(4\pi)^2} \frac{11a(m_\rho)}{96} \ ,
\label{l10}
\end{eqnarray}
where the last term is the finite order correction
from the $\rho$-$\pi$ loop contribution.

The $\rho$-$\pi$-$\pi$ coupling constant $g_{\rho\pi\pi}$:

Strictly speaking, we have to include a higher derivative type
$z_4$ term listed in Ref.~\cite{Tanabashi} 
(see Appendix~\ref{sec:DEH}).
However, a detailed analysis of the model~\cite{HSc}
does not require its existence~\cite{foot:mix}.
Hence we neglect the $z_4$ term.
If we simply read the $\rho$-$\pi$-$\pi$ interaction
from Eq.~(\ref{Lagrangian}), 
we would obtain 
$g_{\rho\pi\pi} = g(m_\rho) F_\sigma^2(m_\rho) / 2 F_\pi^2(m_\rho)$.
However, $g_{\rho\pi\pi}$
should be defined for on-shell $\rho$ and $\pi$'s.
While $F_\sigma^2$ and $g^2$ do not 
run for $\mu<m_\rho$,  $F_\pi^2$ does run.
The on-shell pion decay constant is given by $F_\pi(0)$.  Thus we have
to use $F_\pi(0)$ to define the on-shell $\rho$-$\pi$-$\pi$ coupling
constant.  The resultant expression is given by
\begin{equation}
g_{\rho\pi\pi} = \frac{g(m_\rho)}{2}
\frac{F_\sigma^2(m_\rho)}{F_\pi^2(0)} \ .
\label{g rho pi pi}
\end{equation}

The Gasser-Leutwyler parameter $L_9$~\cite{GL}:

Similarly to the $z_4$-term contribution
to $g_{\rho\pi\pi}$ we neglect the
contribution from the higher derivative type
$z_6$ term~\cite{Tanabashi}.
The resultant relation between $L_9$ and the parameters of the HLS  is
given by~\cite{Tanabashi}
\begin{equation}
L_9(m_\rho) = \frac{1}{4}
\left( \frac{1}{g^2(m_\rho)} - z_3(m_\rho) \right) \ .
\label{l9}
\end{equation}

We further define the parameter $a(0)$ by the direct
$\gamma$-$\pi$-$\pi$ interaction in the second term in 
Eq.~(\ref{Lagrangian}).  
This parameter for on-shell pions is given by
\begin{equation}
a(0) = \frac{F_\sigma^2(m_\rho)}{F_\pi^2(0)} \ ,
\label{a0}
\end{equation}
which should be compared with the parameter $a$ used in the
tree-level analysis, $a=2$ corresponding to the 
vector meson dominance (VMD)~\cite{BKUYY,BKY}.

Then we predict the physical quantities as listed in Table~II.
The predicted values of $g_\rho$, $g_{\rho\pi\pi}$,
$L_9(m_\rho)$ and $L_{10}(m_\rho)$ remarkably
agree with experiment within 10\%,
although $L_{10}(m_\rho)$ is somewhat sensitive to the values of
$\Lambda_{\rm QCD}$ and $\Lambda$~\cite{foot:Borel}.
Moreover, we have $a(0)\simeq2$, although $a(\Lambda)\simeq a(m_\rho)
\simeq1$.

\begin{center}
\vspace{0.2cm}
\small
\begin{tabular}{|c|c||c|c|c|c|c|c|}
\hline
$\Lambda_{\rm QCD}$ 
  & $\Lambda$ & $g_\rho$ & $g_{\rho\pi\pi}$ 
  & $L_9(m_\rho)$ & $L_{10}(m_\rho)$ 
  & $a(0)$ \\
\hline
0.35 & 1.10 & 0.112 &  6.17 & 7.61 & -5.04 & 1.99 \\
     & 1.20 & 0.108 &  6.20 & 7.37 & -4.26 & 2.01 \\
\hline
0.40 & 1.10 & 0.118 &  6.05 & 7.83 & -6.14 & 1.91 \\
     & 1.20 & 0.114 &  6.12 & 7.67 & -5.36 & 1.96 \\
\hline
\multicolumn{2}{|c||}{Exp.} & 0.118$\pm$0.003 & 6.04$\pm$0.04 
  & 6.9$\pm$0.7 & -5.2$\pm$0.3 & \\
\hline
\end{tabular}
\\
\vspace{0.2cm}
\begin{minipage}{8.3cm}
TABLE~II. \ 
Physical quantities predicted by the Wilsonian matching
conditions and the Wilsonian RGE's.
The units of $\Lambda_{\rm QCD}$ and $\Lambda$ are GeV, and
that of $g_\rho$ is GeV$^2$.
Values of $L_9(m_\rho)$ and $L_{10}(m_\rho)$ are scaled by a factor of
$10^3$.
Experimental values of $g_\rho$ and $g_{\rho\pi\pi}$ are derived from 
$\Gamma(\rho \rightarrow e^+ e^-) = (6.77\pm0.32)$\,keV
and 
$\Gamma(\rho^0\rightarrow\pi^+\pi^-) = (150.8\pm2.0)$\,MeV~\cite{PDG},
respectively.  Those of $L_9(m_\rho)$ and $L_{10}(m_\rho)$ are taken
from Ref.~\cite{EGLPR}.
\end{minipage}
\vspace{0.2cm}
\end{center}

Some comments are in order.

The Wilsonian matching 
condition~(\ref{match A}) 
and the input
values of $F_\pi(0)$ and $m_\rho$ 
together with the Wilsonian RGE's 
determine $F_\pi(m_\rho)$, $a(m_\rho)$ and $g(m_\rho)$, and
hence $g_{\rho\pi\pi}$.
The Wilsonian matching condition~(\ref{match V}) 
with the above three parameters
determine $z_3(m_\rho)$, the value
actually needed~\cite{foot:ree} to explain the
experimental value of $g_\rho$.
The value of $z_3(m_\rho)$ together with $g(m_\rho)$ determines
$L_9(m_\rho)$.  Finally, the
Wilsonian matching condition~(\ref{match z})
with the values of $F_\pi(\Lambda)$, $a(\Lambda)$,
$g(\Lambda)$ and $z_3(\Lambda)$ determines
$z_2(m_\rho) - z_1(m_\rho)$, which
gives only a small correction to
$L_{10}(m_\rho)$.  
Although the tree level $\rho$ contribution to
$L_{10}(m_\rho)$ is large,
the finite $\rho$-$\pi$ loop correction cancels a part of it.
The resultant value of
$L_{10}(m_\rho)$ is close to experiment.

The Kawarabayashi-Suzuki-Riazuddin-Fayyazuddin
(KSRF) (I) relation $g_\rho = 2 g_{\rho\pi\pi} F_\pi^2$~\cite{KSRF}
holds as a low energy theorem
of the HLS~\cite{LET,HY,LET:2}.
Here this is satisfied as follows:
In the low energy limit higher derivative terms like $z_3$
do not contribute, and
the $\rho$-$\gamma$ mixing strength becomes $g_\rho(0) = g(m_\rho)
F_\sigma^2(m_\rho)$. 
Comparing this with $g_{\rho\pi\pi}$ in 
Eq.~(\ref{g rho pi pi})~\cite{foot:rpp},
we can easily read that the low energy theorem is satisfied.  If we use
the experimental values, the KSRF (I) relation is violated by about
10\%.  As discussed above, this deviation is explained by the
existence of the $z_3$ term.

The KSRF (II) relation
$m_\rho^2 = 2 g_{\rho\pi\pi}^2 F_\pi^2$~\cite{KSRF}
is approximately satisfied by
the on-shell quantities even though $a(m_\rho) \simeq 1$.
This is seen as follows.
Equation~(\ref{g rho pi pi}) with Eq.~(\ref{a0}) and $m_\rho^2 =
g^2(m_\rho) F_\sigma^2(m_\rho)$ leads to $2 g_{\rho\pi\pi}^2
F_\pi^2(0) = m_\rho^2 \left( a(0)/2 \right)$.  Thus $a(0) \simeq 2$
leads to the approximate KSRF (II) relation.
Furthermore, $a(0)\simeq2$
implies that the direct $\gamma$-$\pi$-$\pi$ coupling is suppressed
(VMD).

Inclusion of the quadratic divergences into the RGE's 
was essential in the present analysis.
{\it The RGE's with logarithmic divergence alone
would not be consistent with the matching to QCD.}
The bare parameter $F_\pi(\Lambda)=158$\,MeV listed in
Table~I, which is derived by the matching condition
(\ref{match A}), is about double of the physical value
$F_\pi(0)=88$\,MeV.
The logarithmic running by the first term of 
Eq.~(\ref{RGE for Fpi2}) is not enough to change the value of $F_\pi$.
Actually, the present procedure with logarithmic running
would lead to
$g_\rho = 0.11$\,GeV$^2$, $g_{\rho\pi\pi}=10$,
$L_9(m_\rho)=13\times10^{-3}$ and $L_{10}(m_\rho)=+4.5\times10^{-3}$.
The latter three badly disagree with experiment~\cite{foot:log}.

\section{Discussion}

It is interesting to apply the Wilsonian matching proposed in this
paper for an analysis of large $N_f$ QCD done in 
Ref.~\cite{HY:letter}.  There it was assumed that the
ratio $F_\pi^2(\Lambda)/\Lambda^2$ has a small $N_f$ dependence.
As is easily read from Eq.~(\ref{match A}), the Wilsonian matching
condition implies that the ratio actually has a small $N_f$
dependence.  The analysis of the large $N_f$ chiral restoration of QCD
in this line will be done in a separate paper~\cite{HY:VM}.

\section*{Acknowledgments}

We would like to thank Howard Georgi and Volodya Miransky for useful
discussions.
K.Y. thanks Howard Georgi for hospitality during his stay at Harvard
where a part of this work was done.
This work is supported in part by Grant-in-Aid for Scientific Research
(B)\#11695030 (K.Y.), (A)\#12014206 (K.Y.) and (A)\#12740144 (M.H.),
and by the Oversea Research Scholar Program of the Ministry of
Education, Science, Sports and Culture (K.Y.).

\appendix

\section{Derivative Expansion in HLS}
\label{sec:DEH}

In chiral perturbation theory (ChPT)~\cite{Weinberg,GL}
the derivative expansion is systematically done by 
using the fact that the pseudoscalar meson masses are small compared
with the chiral symmetry breaking scale $\Lambda_\chi$.
The chiral symmetry breaking scale is considered as the scale where
the derivative expansion breaks down.
{}From the naive dimensional analysis~\cite{MG} $\Lambda_\chi$ is
estimated as
\begin{equation}
\Lambda_\chi \simeq 4 \pi F_\pi \sim 1.1 \, \mbox{GeV} \ ,
\end{equation}
which also agrees with the matching scale
(\ref{matching scale}) used in the text.
Since the $\rho$ meson and its flavor partners are lighter than this
scale, one may consider that a derivative expansion with including
vector mesons is possible.
Actually, 
the first one-loop calculation based on this notion was done in
Ref.~\cite{HY}.  
There it was shown that the low energy theorem of the 
HLS~\cite{LET} holds at one loop.
This low energy theorem was proved to hold at any loop order
in Ref.~\cite{LET:2}.
Moreover, a systematic counting scheme in the framework of the HLS
was proposed in Ref.~\cite{Tanabashi}.
A key point there was the fact that the vector meson masses in
the HLS become small in the limit of the small HLS gauge coupling.
It turns out that 
such a limit can actually be realized in QCD
when the massless flavor $N_f$
becomes large as was demonstrated in Refs.~\cite{HY:letter,HY:VM}.
Then one can perform the derivative expansion with including the
vector mesons in the idealized world where the vector meson masses are
small and extrapolate the results to the world where the vector meson
masses take the experimental values.
Although the expansion parameter is not very small,
\begin{equation}
\frac{m_\rho^2}{(4\pi F_\pi)^2} \sim 0.4 \ ,
\label{expar}
\end{equation}
that procedure seems to work in the real world. (See, e.g., the
discussion in Ref.~\cite{LET:2}.)
Here we apply such a systematic expansion to the realistic case
$N_f=3$.

For the complete analysis at one loop, we need to include the term
having external scalar and pseudoscalar source fields $S$ and
$P$, as shown in Ref.~\cite{Tanabashi}.
These are included through the external source field $\hat{\chi}$
defined by
\begin{eqnarray}
&&
\widehat{\chi} \equiv \xi_{\rm L} \, \chi \, \xi_{\rm R}^{\dag} \ ,
\label{def:chihat}
\\
&& \chi \equiv 2 B \left( S + i P \right)
\ ,
\label{def:chi}
\end{eqnarray}
where $B$ is a constant parameter.
If there is an explicit chiral symmetry breaking due to the current
quark mass, it is introduced as the vacuum expectation value (VEV) of
the external scalar source field:
\begin{equation}
\langle S \rangle = {\cal M} =
\left( \begin{array}{ccc}
m_1 & & \\
 & \ddots & \\
 & & m_{N_f} \\
\end{array} \right) \ .
\label{quark mass matrix}
\end{equation}
However, in the present paper, we work in the
chiral limit, so that we take the VEV to zero.

Now, let us summarize the counting rule of the present analysis.
As in the ChPT in Ref.~\cite{GL},
the derivative and the external gauge fields ${\cal L}_\mu$ and 
${\cal R}_\mu$ are counted as ${\cal O}(p)$, while the 
external source fields $\widehat{\chi}$ (or $\chi$) is
counted as ${\cal O}(p^2)$ since the VEV of $\widehat{\chi}$ is the
square of the pseudoscalar meson mass, $\langle\widehat{\chi}\rangle =
m_\pi^2$:
\begin{eqnarray}
&& \partial_\mu \sim {\cal L}_\mu \sim 
  {\cal R}_\mu \sim {\cal O}(p) \ , \nonumber\\
&& \widehat{\chi} \sim {\cal O}(p^2) \ .
\end{eqnarray}
For consistency of the covariant derivative shown in
Eq.~(\ref{covder}) we assign ${\cal O}(p)$ to 
$V_\mu \equiv g \rho_\mu$:
\begin{equation}
V_\mu = g \rho_\mu \sim {\cal O}(p) \ .
\label{V:order}
\end{equation}
The above counting rules are the same as those in the ChPT.
An essential difference between the order counting in the HLS and
that in the ChPT is in the counting rule for the vector meson
mass.
In an extension of the ChPT (see, e.g., Ref.~\cite{EGLPR})
the vector meson mass is counted as
${\cal O}(1)$ at the scale below the vector meson mass.
However, as discussed around Eq.~(\ref{expar}), we are performing the
derivative expansion in the HLS by regarding the vector meson
as light.
Thus, similarly to the square of the pseudoscalar meson mass,
we assign ${\cal O}(p^2)$ to the square of the vector meson mass:
\begin{equation}
m_\rho^2 = g^2 F_\sigma^2 \sim {\cal O}(p^2) \ .
\end{equation}
Since the vector meson mass becomes small in the limit of small
HLS gauge coupling,
we should assign ${\cal O}(p)$ to the
HLS gauge coupling $g$, not to $F_\sigma$:
\begin{equation}
g \sim {\cal O}(p) \ .
\label{g:order}
\end{equation}
This is the most important part in the counting rules in the HLS.
By comparing the order for $g$ in Eq.~(\ref{g:order}) with that for
$g \rho_\mu$ in Eq.~(\ref{V:order}), the $\rho_\mu$ field should be
counted as ${\cal O}(1)$.
Then the kinetic term of the HLS gauge boson is counted as 
${\cal O}(p^2)$ which is of the same order as the kinetic term of the
pseudoscalar meson.

With the above counting rules the leading order Lagrangian is given
by~\cite{BKUYY,BKY,Tanabashi}
\begin{eqnarray}
{\cal L}_{(2)} &=& F_\pi^2 \, \mbox{tr} 
\left[ \hat{\alpha}_{\perp\mu} \hat{\alpha}_{\perp}^\mu \right]
+ F_\sigma^2 \, \mbox{tr}
\left[ 
  \hat{\alpha}_{\parallel\mu} \hat{\alpha}_{\parallel}^\mu
\right]
\nonumber\\
&& 
- \frac{1}{2g^2} \, \mbox{tr} 
\left[ V_{\mu\nu} V^{\mu\nu} \right] 
+ \frac{1}{4} F_\chi^2 \mbox{tr}
\left[ \hat{\chi} + \hat{\chi}^\dag \right ]
\ ,
\label{leading Lagrangian}
\end{eqnarray}
where as discussed above
we rescaled the vector meson field as
\begin{equation}
V_\mu = g \rho_\mu \ .
\end{equation}
$F_\chi$ in the fourth term in Eq.~(\ref{leading Lagrangian}),
which was absent in the previous analysis done in
Ref.~\cite{Tanabashi},
was introduced to renormalize the quadratically divergent
correction to the fourth term.
We note that this $F_\chi$ agrees with $F_\pi$ at the tree level.
In the present analysis we will not consider the renormalization
effect of $F_\chi$.

A complete list of the ${\cal O}(p^4)$ Lagrangian for 
the SU($N_f$) case
is shown in Ref.~\cite{Tanabashi}, where use was made of
the equations of motion
\begin{eqnarray}
&&
D_\mu \hat{\alpha}_{\perp}^\mu =
- i \left( a- 1 \right)
\left[
  \hat{\alpha}_{\parallel\mu} \,,\,
  \hat{\alpha}_\perp^\mu
\right]
\nonumber\\
&& \quad
{}+ \frac{i}{4} \frac{F_\chi^2}{F_\pi^2} 
\left(
  \hat{\chi} - \hat{\chi}^\dag
  - \frac{1}{N_f} \, \mbox{tr}
  \left[ \hat{\chi} - \hat{\chi}^\dag \right]
\right)
+ {\cal O}(p^4)
\ ,
\label{EOM Npi2}
\\
&&
D_\mu \hat{\alpha}_{\parallel}^\mu 
= {\cal O}(p^4)
\ ,
\label{EOM Nsig2}
\\
&&
D_\nu {V}^{\nu\mu}
=
g^2 f_\sigma^2 
\hat{\alpha}_{\parallel}^\mu 
+ {\cal O}(p^4)
\ ,
\label{EOM Nvec2}
\end{eqnarray}
and the identities
\begin{eqnarray}
&&
D_\mu \hat{\alpha}_{\perp\nu}
- D_\nu \hat{\alpha}_{\perp\mu}
\nonumber\\
&& \quad
=
i \left[ 
  \hat{\alpha}_{\parallel\mu} \,,\, \hat{\alpha}_{\perp\nu}
\right]
+
i \left[ 
  \hat{\alpha}_{\perp\mu} \,,\, \hat{\alpha}_{\parallel\nu}
\right]
- \widehat{\cal A}_{\mu\nu}
\ ,
\label{rel:perp}
\\
&&
D_\mu \hat{\alpha}_{\parallel\nu}
- D_\nu \hat{\alpha}_{\parallel\mu}
\nonumber\\
&& \quad
=
i \left[ 
  \hat{\alpha}_{\parallel\mu} \,,\, \hat{\alpha}_{\parallel\nu}
\right]
+
i \left[ 
  \hat{\alpha}_{\perp\mu} \,,\, \hat{\alpha}_{\perp\nu}
\right]
+ \widehat{\cal V}_{\mu\nu}
- V_{\mu\nu}
\ .
\label{rel:parallel}
\end{eqnarray}
Below we write the ${\cal O}(p^4)$ terms listed in
Ref.~\cite{Tanabashi} for the reader's convenience:
\begin{eqnarray}
&&
{\cal L}_{(4)y} =
y_1 \, \mbox{tr} \left[ 
  \hat{\alpha}_{\perp\mu} \hat{\alpha}_\perp^\mu
  \hat{\alpha}_{\perp\nu} \hat{\alpha}_\perp^\nu
\right]
+ y_2 \, \mbox{tr} \left[
  \hat{\alpha}_{\perp\mu} \hat{\alpha}_{\perp\nu}
  \hat{\alpha}^\mu_\perp \hat{\alpha}^\nu_\perp
\right]
\nonumber\\
&&\ 
{}+ y_3 \, \mbox{tr} \left[
  \hat{\alpha}_{\parallel\mu} \hat{\alpha}_\parallel^\mu
  \hat{\alpha}_{\parallel\nu} \hat{\alpha}_\parallel^\nu
\right]
+ y_4 \, \mbox{tr} \left[
  \hat{\alpha}_{\parallel\mu} \hat{\alpha}_{\parallel\nu}
  \hat{\alpha}^\mu_\parallel \hat{\alpha}^\nu_\parallel
\right]
\nonumber\\
&&\ 
{}+ y_5 \, \mbox{tr} \left[
  \hat{\alpha}_{\perp\mu} \hat{\alpha}_\perp^\mu
  \hat{\alpha}_{\parallel\nu} \hat{\alpha}_\parallel^\nu
\right]
+ y_6 \, \mbox{tr} \left[
  \hat{\alpha}_{\perp\mu} \hat{\alpha}_{\perp\nu}
  \hat{\alpha}^\mu_\parallel \hat{\alpha}^\nu_\parallel
\right]
\nonumber\\
&&\ 
{}+ y_7 \, \mbox{tr} \left[
  \hat{\alpha}_{\perp\mu} \hat{\alpha}_{\perp\nu}
  \hat{\alpha}^\nu_\parallel \hat{\alpha}^\mu_\parallel
\right]
\nonumber\\
&&\ 
{}+ y_8 \, \left\{
\mbox{tr} \left[ 
  \hat{\alpha}_{\perp\mu} \hat{\alpha}_\parallel^\mu
  \hat{\alpha}_{\perp\nu} \hat{\alpha}_\parallel^\nu
\right]
+ \mbox{tr} \left[
  \hat{\alpha}_{\perp\mu} \hat{\alpha}_{\parallel\nu} 
  \hat{\alpha}_\perp^\nu \hat{\alpha}_\parallel^\mu
\right] \right\}
\nonumber\\
&&\ 
{}+ y_9 \, \mbox{tr} \left[
  \hat{\alpha}_{\perp\mu} \hat{\alpha}_{\parallel\nu}
  \hat{\alpha}^\mu_\perp \hat{\alpha}^\nu_\parallel
\right]
\nonumber\\
&&\ 
{}+ y_{10} \left(
\mbox{tr} \left[
  \hat{\alpha}_{\perp\mu} \hat{\alpha}_\perp^\mu
\right] \right)^2
+ y_{11} \, \mbox{tr} \left[
  \hat{\alpha}_{\perp\mu} \hat{\alpha}_{\perp\nu}
\right] 
\mbox{tr} \left[
  \hat{\alpha}^\mu_\perp \hat{\alpha}^\nu_\perp
\right] 
\nonumber\\
&&\ 
{}+ y_{12} \left( \mbox{tr} \left[ 
  \hat{\alpha}_{\parallel\mu} \hat{\alpha}_\parallel^\mu
\right] \right)^2
+ y_{13} \, \mbox{tr} \left[
  \hat{\alpha}_{\parallel\mu} \hat{\alpha}_{\parallel\nu}
\right]
\mbox{tr} \left[
  \hat{\alpha}^\mu_\parallel \hat{\alpha}^\nu_\parallel
\right]
\nonumber\\
&&\ 
{}+ y_{14} \, \mbox{tr} \left[
  \hat{\alpha}_{\perp\mu} \hat{\alpha}_\perp^\mu
\right]
\mbox{tr} \left[
  \hat{\alpha}_{\parallel\nu} \hat{\alpha}_\parallel^\nu
\right]
\nonumber\\
&&\ 
{}+ y_{15} \, \mbox{tr} \left[
  \hat{\alpha}_{\perp\mu} \hat{\alpha}_{\perp\nu}
\right]
\mbox{tr} \left[
  \hat{\alpha}^\mu_\parallel \hat{\alpha}^\nu_\parallel
\right]
\nonumber\\
&&\ 
{}+ y_{16} \left( \mbox{tr} \left[
  \hat{\alpha}_{\perp\mu} \hat{\alpha}_\parallel^\mu
\right] \right)^2
+ y_{17} \, \mbox{tr} \left[
  \hat{\alpha}_{\perp\mu} \hat{\alpha}_{\parallel\nu}
\right]
\mbox{tr} \left[
  \hat{\alpha}^\mu_\perp \hat{\alpha}^\nu_\parallel
\right]
\nonumber\\
&&\ 
{}+ y_{18} \, \mbox{tr} \left[
  \hat{\alpha}_{\perp\mu} \hat{\alpha}_{\parallel\nu}
\right]
\mbox{tr} \left[
  \hat{\alpha}^\mu_\parallel \hat{\alpha}^\nu_\perp 
\right]
\ ,
\nonumber
\\
&&
{\cal L}_{(4)w} =
w_1 \, \frac{F_\chi^2}{F_\pi^2} \, \mbox{tr} \left[
  \hat{\alpha}_{\perp\mu} \hat{\alpha}_\perp^\mu
  \left( \hat{\chi} + \hat{\chi}^\dag \right)
\right]
\nonumber\\
&&\ 
{}+ w_2 \, \frac{F_\chi^2}{F_\pi^2} \, \mbox{tr} \left[
  \hat{\alpha}_{\perp\mu} \hat{\alpha}_\perp^\mu
\right]
\mbox{tr} \left[
  \hat{\chi} + \hat{\chi}^\dag
\right]
\nonumber\\
&&\ 
{}+ w_3 \, \frac{F_\chi^2}{F_\pi^2} \, \mbox{tr} \left[
  \hat{\alpha}_{\parallel\mu} \hat{\alpha}_\parallel^\mu
  \left( \hat{\chi} + \hat{\chi}^\dag \right)
\right]
\nonumber\\
&&\ 
{}+ w_4 \, \frac{F_\chi^2}{F_\pi^2} \, \mbox{tr} \left[
  \hat{\alpha}_{\parallel\mu} \hat{\alpha}_\parallel^\mu
\right]
\mbox{tr} \left[
  \hat{\chi} + \hat{\chi}^\dag
\right]
\nonumber\\
&&\ 
{}+ w_5 \, \frac{F_\chi^2}{F_\pi^2} \, \mbox{tr} \left[ 
  \left( 
    \hat{\alpha}_{\perp\mu} \hat{\alpha}_\parallel^\mu
    - \hat{\alpha}_\parallel^\mu \hat{\alpha}_{\perp\mu}
  \right)
  \left( \hat{\chi} - \hat{\chi}^\dag \right)
\right]
\nonumber\\
&&\ 
{}+ w_6 \, \frac{F_\chi^4}{F_\pi^4} \, \mbox{tr} \left[
  \left( \hat{\chi} + \hat{\chi}^\dag \right)^2
\right]
+ w_7 \, \frac{F_\chi^4}{F_\pi^4} \left( \mbox{tr} \left[
  \hat{\chi} + \hat{\chi}^\dag
\right] \right)^2
\nonumber\\
&&\ 
{}+ w_8 \, \frac{F_\chi^4}{F_\pi^4} \, \mbox{tr} \left[
  \left( \hat{\chi} - \hat{\chi}^\dag \right)^2
\right]
+ w_9 \, \frac{F_\chi^4}{F_\pi^4} \left( \mbox{tr} \left[
  \hat{\chi} - \hat{\chi}^\dag
\right] \right)^2
\ ,
\nonumber
\\
&&
{\cal L}_{(4)z} =
z_1 \,\mbox{tr}\left[ \hat{\cal V}_{\mu\nu} \hat{\cal V}^{\mu\nu} \right]
+ z_2 \,\mbox{tr}\left[ \hat{\cal A}_{\mu\nu} \hat{\cal A}^{\mu\nu} \right]
\nonumber\\
&&\ 
{}+ z_3 \,\mbox{tr}\left[ \hat{\cal V}_{\mu\nu} V^{\mu\nu} \right]
\nonumber\\
&&\ 
{} + i z_4 \,\mbox{tr}\left[ 
  V_{\mu\nu} \hat{\alpha}_\perp^\mu \hat{\alpha}_\perp^\nu 
\right]
+ i z_5 \,\mbox{tr}\left[ 
  V_{\mu\nu} \hat{\alpha}_\parallel^\mu \hat{\alpha}_\parallel^\nu 
\right]
\nonumber\\
&&\ 
{} + i z_6 \,\mbox{tr}\left[ 
  \hat{\cal V}_{\mu\nu} \hat{\alpha}_\perp^\mu \hat{\alpha}_\perp^\nu 
\right]
+ i z_7 \,\mbox{tr}\left[ 
  \hat{\cal V}_{\mu\nu} \hat{\alpha}_\parallel^\mu \hat{\alpha}_\parallel^\nu 
\right]
\nonumber\\
&&\ 
{} + i z_8 \,\mbox{tr}\left[ 
  \hat{\cal A}_{\mu\nu} 
  \left( \hat{\alpha}_\perp^\mu \hat{\alpha}_\parallel^\nu 
         + \hat{\alpha}_\parallel^\mu \hat{\alpha}_\perp^\nu \right)
\right]
\ .
\label{Lag: ywz terms}
\end{eqnarray}
We note here that among those given in Eq.~(\ref{Lag: ywz terms})
only $z_1$, $z_2$ and $z_3$ are relevant to the present analysis
which is confined to the two-point functions in the chiral symmetric
limit.

In section~\ref{sec:DBPHL} we discussed the low energy parameters 
$L_9$ and $L_{10}$ of the ChPT defined in Ref.~\cite{GL}.
Below we shall list the ${\cal O}(p^4)$ terms in the ChPT for 
the reader's convenience:
\begin{eqnarray}
&&
{\cal L}_{(4)}^{\rm ChPT} =
L_1\, \left( \mbox{tr}
  \left[ \nabla_\mu U^\dag \nabla^\mu U \right]
  \right)^2
\nonumber\\
&& \quad
{}+ L_2\,
\mbox{tr} \left[ \nabla_\mu U^\dag \nabla_\nu U \right]
  \mbox{tr}\, \left[ \nabla^\mu U^\dag \nabla^\nu U \right]
\nonumber\\
&& \quad
{}+ L_3\,
\mbox{tr}
  \left[ \nabla_\mu U^\dag \nabla^\mu U 
  \nabla_\nu U^\dag \nabla^\nu U \right]
\nonumber\\
&& \quad
{}+ L_4\,
\mbox{tr} \left[ \nabla_\mu U^\dag \nabla^\mu U \right]
  \mbox{tr}\, \left[ \chi^\dag U + \chi U^\dag \right ]
\nonumber\\
&& \quad
{}+ L_5\,
\mbox{tr} \left[ \nabla_\mu U^\dag \nabla^\mu U 
  \left(\chi^\dag U + U^\dag \chi \right) \right ]
\nonumber\\
&& \quad
{}+ L_6\,
\left( \mbox{tr} \left[ \chi^\dag U + \chi U^\dag \right ] \right)^2
\nonumber\\
&& \quad
{}+ L_7\,
\left( \mbox{tr} \left[ \chi^\dag U - \chi U^\dag \right ] \right)^2
\nonumber\\
&& \quad
{}+ L_8\,
\mbox{tr} \left[ 
  \chi^\dag U \chi^\dag U + \chi U^\dag \chi U^\dag \right ]
\nonumber\\
&& \quad
{}- i\, L_9\,
\mbox{tr} \left[
{\cal L}_{\mu\nu} \nabla^\mu U \nabla^\nu U^\dag
+ {\cal R}_{\mu\nu} \nabla^\mu U^\dag \nabla^\nu U
\right]
\nonumber\\
&& \quad
{}+ L_{10}\,
\mbox{tr} \left[
U^\dag {\cal L}_{\mu\nu} U {\cal R}_{\mu\nu} \right]
\nonumber\\
&& \quad
{}+ H_1\,
\mbox{tr} \left[ 
{\cal L}_{\mu\nu} {\cal L}^{\mu\nu} +
{\cal R}_{\mu\nu} {\cal R}^{\mu\nu}
\right]
\nonumber\\
&& \quad
{}+ H_2\,
\mbox{tr} \left[ \chi^\dag \chi \right]
\ ,
\label{p4:ChPT}
\end{eqnarray}
where ${\cal L}_{\mu\nu}$ and ${\cal R}_{\mu\nu}$ are the field
strengths of the external gauge fields ${\cal L}_\mu$ and
${\cal R}_\mu$, respectively, $\chi$ is defined in 
Eq.~(\ref{def:chi}), and $U$ is defined as
[see Eq.~(\ref{xiLR})]
\begin{eqnarray}
&& U \equiv e^{2i \pi/F_\pi} = \xi_{\rm L}^\dag \xi_{\rm R} \ .
\label{def:U}
\end{eqnarray}
The covariant derivative acting on $U$ is defined as
[see Eq.~(\ref{covder})]
\begin{equation}
\nabla_\mu U \equiv \partial_\mu - i {\cal L}_\mu U + i U {\cal R}_\mu
\ .
\end{equation}
Here we note that the above expression in Eq.~(\ref{p4:ChPT})
is valid for $N_f=3$,
and for $N_f\geq4$ there is an extra term given by
\begin{eqnarray}
\mbox{tr}
  \left[ \nabla_\mu U \nabla_\nu U^\dag
  \nabla^\mu U \nabla^\nu U^\dag \right]
\ .
\end{eqnarray}

The relations at the tree level between the parameters in the ChPT and 
those in the HLS are obtained by integrating out the $\rho$ field
with the vector meson mass regarded as ${\cal O}(1)$.
[This implies that the HLS gauge coupling $g$ is regarded as 
${\cal O}(1)$.]
In this case the equation of motion (\ref{EOM Nvec2}) leads to
\begin{equation}
\hat{\alpha}_{\parallel}^\mu = \frac{1}{m_\rho^2} {\cal O}(p^3)
\label{para0}
\end{equation}
and, thus,
\begin{equation}
V_{\mu\nu} = \hat{\cal V}_{\mu\nu} + i 
\left[ \hat{\alpha}_{\perp\mu} \,,\, \hat{\alpha}_{\perp\nu} \right]
+ \frac{1}{m_\rho^2} {\cal O}(p^4) \ .
\label{V0}
\end{equation}
Furthermore, we have
\begin{equation}
\hat{\alpha}_{\perp\mu} 
= \frac{i}{2} \xi_{\rm L} \cdot \nabla U \cdot \xi_{\rm R}^\dag
= \frac{1}{2i} \xi_{\rm R} \cdot \nabla U^\dag \cdot \xi_{\rm L}^\dag
\ 
\label{perpU}
\end{equation}
and
\begin{eqnarray}
\mbox{tr} \left[
  \hat{\cal V}_{\mu\nu} \hat{\cal V}^{\mu\nu}
\right]
&=& \frac{1}{4}
\mbox{tr} \left[ 
{\cal L}_{\mu\nu} {\cal L}^{\mu\nu} +
{\cal R}_{\mu\nu} {\cal R}^{\mu\nu}
\right]
\nonumber\\
&& \quad
{}- \frac{1}{2}
\mbox{tr} \left[
U^\dag {\cal L}_{\mu\nu} U {\cal R}_{\mu\nu} \right]
\ ,
\nonumber\\
\mbox{tr}\left[ \hat{\cal A}_{\mu\nu} \hat{\cal A}^{\mu\nu} \right]
&=& \frac{1}{4}
\mbox{tr} \left[ 
{\cal L}_{\mu\nu} {\cal L}^{\mu\nu} +
{\cal R}_{\mu\nu} {\cal R}^{\mu\nu}
\right]
\nonumber\\
&& \quad
{}+ \frac{1}{2}
\mbox{tr} \left[
U^\dag {\cal L}_{\mu\nu} U {\cal R}_{\mu\nu} \right]
\ ,
\end{eqnarray}
where we used Eq.~(\ref{def:AV}) with Eq.~(\ref{def:U}).
By substituting Eq.~(\ref{perpU}) 
into the HLS Lagrangian,
the first and fourth terms in the leading order HLS Lagrangian 
(\ref{leading Lagrangian}) become the leading order ChPT Lagrangian:
\begin{equation}
{\cal L}_{(2)}^{\rm ChPT} =
\frac{F_\pi^2}{4} \mbox{tr}
\left[ \nabla_\mu U^\dag \nabla^\mu U \right]
+ \frac{F_\pi^2}{4} \mbox{tr}
\left[ \chi U^\dag + \chi^\dag U \right ]
\ ,
\label{leading ChPT}
\end{equation}
where we took $F_\chi=F_\pi$.
In addition, 
the second term in Eq.~(\ref{leading Lagrangian})
with Eq.~(\ref{para0}) substituted
becomes of ${\cal O}(p^6)$ in the ChPT
and the third term (the kinetic term of the HLS gauge boson) 
with Eq.~(\ref{V0}) becomes
of ${\cal O}(p^4)$ in the ChPT.
In the ${\cal O}(p^4)$ HLS Lagrangian (\ref{Lag: ywz terms})
the terms including 
$\hat{\alpha}_{\parallel}^\mu$ become of higher order in the ChPT.
The remaining terms together with the kinetic term of
the HLS gauge boson [the third term in Eq.~(\ref{leading Lagrangian})]
become the ${\cal O}(p^4)$ ChPT Lagrangian.
Below, we list the correspondence between the parameters in the
HLS and the ${\cal O}(p^4)$ ChPT parameters at 
the tree level for $N_f=3$:
\begin{eqnarray}
&& 
L_1 \mathop{\Longleftrightarrow}_{tree}
\frac{1}{32 g^2} + \frac{1}{32} y_2 + \frac{1}{16} y_{10}
\ ,
\nonumber\\
&& 
L_2 \mathop{\Longleftrightarrow}_{tree}
\frac{1}{16 g^2} + \frac{1}{16} y_2 + \frac{1}{16} y_{11}
\ ,
\nonumber\\
&& 
L_3 \mathop{\Longleftrightarrow}_{tree}
- \frac{3}{16 g^2} + \frac{1}{16} y_1 - \frac{1}{8} y_2
\ ,
\nonumber\\
&& 
L_4 \mathop{\Longleftrightarrow}_{tree} \frac{1}{4} w_2
\ ,
\nonumber\\
&& 
L_5 \mathop{\Longleftrightarrow}_{tree} \frac{1}{4} w_1
\ ,
\nonumber\\
&& 
L_6 \mathop{\Longleftrightarrow}_{tree} w_7
\ ,
\nonumber\\
&& 
L_7 \mathop{\Longleftrightarrow}_{tree} w_9
\ ,
\nonumber\\
&& 
L_8 \mathop{\Longleftrightarrow}_{tree} \left( w_6 + w_8 \right)
\ ,
\nonumber\\
&& 
L_9 \mathop{\Longleftrightarrow}_{tree} 
\frac{1}{4} \left( \frac{1}{g^2} - z_3 \right)
+ \frac{1}{8} \left( z_4 + z_6 \right)
\ ,
\nonumber\\
&& 
L_{10} \mathop{\Longleftrightarrow}_{tree} 
- \frac{1}{4g^2} + \frac{1}{2} \left( z_3 - z_2 + z_1 \right)
\ ,
\nonumber\\
&& 
H_1 \mathop{\Longleftrightarrow}_{tree} 
- \frac{1}{8g^2} + \frac{1}{4} \left( z_3 + z_2 + z_1 \right)
\ ,
\nonumber\\
&& 
H_2 \mathop{\Longleftrightarrow}_{tree} 2 \left( w_6 - w_8 \right)
\ ,
\end{eqnarray}
where we took $F_\chi = F_\pi$.
It should be noticed that the above relations are valid at the tree
level.
As discussed in Ref.~\cite{Tanabashi} we have to relate these
at the one-loop level where
finite order corrections appear in several relations:
The relation between $L_{10}$ and the parameters in the HLS
becomes Eq.~(\ref{l10}) by adding finite order corrections.
[We will derive this finite order correction later in
Eq.~(\ref{l10fin}).]
On the other hand, there is no substantial finite order correction to
the relation for $L_9$.  Moreover, as discussed above Eq.~(\ref{g rho
pi pi}) a detailed analysis~\cite{HSc}
using a similar model~\cite{KS}
does not require the existence of 
a higher derivative type
$z_4$ term as well as a $z_6$ term.
Hence we neglected the $z_4$ and $z_6$ terms and obtained the relation
in Eq.~(\ref{l9}).

\section{Background Gauge Field Method}
\label{sec:BGFM}

We adopt the background gauge field method
to obtain quantum corrections to the parameters.
(For calculations in other gauges, see Ref.~\cite{HY} for the
$R_\xi$-like gauge and Ref.~\cite{LET:2} for the covariant gauge.)
This appendix is a preparation to calculate the
renormalization group equations in Appendix~\ref{sec:RGE}.
The background field method was used in the ChPT in Ref.~\cite{GL},
and was applied to the HLS in Ref.~\cite{Tanabashi}.
Following Ref.~\cite{Tanabashi}
we introduce the background fields $\overline{\xi}_{\rm L}$
and $\overline{\xi}_{\rm R}$ as
\begin{equation}
\xi_{\rm L,R} = \widehat{\xi}_{\rm L,R} \overline{\xi}_{\rm L,R} \ ,
\end{equation}
where $\hat{\xi}_{\rm L,R}$ denote the quantum fields.
It is convenient to write
\begin{eqnarray}
&&
\widehat{\xi}_{\rm L} = 
\widehat{\xi}_{\rm S} \cdot
\widehat{\xi}_{\rm P}^\dag \ ,
\qquad
\widehat{\xi}_{\rm R} = 
\widehat{\xi}_{\rm S} \cdot
\widehat{\xi}_{\rm P} \ ,
\nonumber\\
&& \quad
\widehat{\xi}_{\rm P} =
\exp \left[ i\, \widehat{\varphi}_\pi^a T_a \right] \ ,
\quad
\widehat{\xi}_{\rm S} =
\exp \left[ i \, \widehat{\varphi}_\sigma^a T_a \right] \ ,
\end{eqnarray}
with $\widehat{\varphi}_\pi$ and $\widehat{\varphi}_\sigma$ being the
quantum fields corresponding to the NG boson $\pi$ and the would-be NG
boson $\sigma$. 
The background field $\overline{V}_\mu$ and the quantum field
$\widehat{v}_\mu$ of the HLS gauge boson are introduced as
\begin{eqnarray}
&&
V_\mu = \overline{V}_\mu + g \widehat{v}_\mu \ .
\end{eqnarray}
We use the following notation for the background fields
including $\overline{\xi}_{\rm L,R}$:
\begin{eqnarray}
\overline{\cal A}_\mu &\equiv&
\frac{1}{2i} \left[
  \partial_\mu \overline{\xi}_{\rm L} \cdot 
    \overline{\xi}_{\rm L}^\dag
  - \partial_\mu \overline{\xi}_{\rm R} \cdot 
    \overline{\xi}_{\rm R}^\dag
\right]
\nonumber\\
&&\ 
+ \frac{1}{2} \left[
  \overline{\xi}_{\rm L} {\cal L}_\mu 
    \overline{\xi}_{\rm L}^\dag
  - \overline{\xi}_{\rm R} {\cal R}_\mu 
    \overline{\xi}_{\rm R}^\dag
\right]
\ ,
\nonumber\\
\overline{\cal V}_\mu &\equiv&
\frac{1}{2i} \left[
  \partial_\mu \overline{\xi}_{\rm L} \cdot 
    \overline{\xi}_{\rm L}^\dag
  + \partial_\mu \overline{\xi}_{\rm R} \cdot 
    \overline{\xi}_{\rm R}^\dag
\right]
\nonumber\\
&&\ 
+ \frac{1}{2} \left[
  \overline{\xi}_{\rm L} {\cal L}_\mu 
    \overline{\xi}_{\rm L}^\dag
  + \overline{\xi}_{\rm R} {\cal R}_\mu 
    \overline{\xi}_{\rm R}^\dag
\right]
\ ,
\end{eqnarray}
which correspond to $\hat{\alpha}_{\perp\mu}$ and 
$\hat{\alpha}_{\parallel\mu} + V_\mu$, respectively.
The field strengths of $\overline{\cal A}_\mu$ and 
$\overline{\cal V}_\mu$ are
defined as
\begin{eqnarray}
\overline{\cal V}_{\mu\nu} &=&
\partial_\mu \overline{\cal V}_\nu 
- \partial_\nu \overline{\cal V}_\mu
- i \left[ \overline{\cal V}_\mu \,,\, \overline{\cal V}_\nu \right]
- i \left[ \overline{\cal A}_\mu \,,\, \overline{\cal A}_\nu \right]
\ ,
\nonumber\\
\overline{\cal A}_{\mu\nu} &=&
\partial_\mu \overline{\cal A}_\nu 
- \partial_\nu \overline{\cal A}_\mu
- i \left[ \overline{\cal V}_\mu \,,\, \overline{\cal A}_\nu \right]
- i \left[ \overline{\cal A}_\mu \,,\, \overline{\cal V}_\nu \right]
\ ,
\end{eqnarray}
which correspond to $\hat{V}_{\mu\nu}$ and $\hat{A}_{\mu\nu}$,
respectively.
In addition we use $\overline{\chi}$ for the background field
corresponding to $\hat{\chi}$:
\begin{equation}
\overline{\chi} \equiv 2 B \overline{\xi}_{\rm L}
\left( S + i P \right) \overline{\xi}_{\rm R}^\dag
\ .
\end{equation}

It should be noticed that the quantum fields as well as the background
fields $\overline{\xi}_{\rm R,L}$ transform homogeneously
under the background gauge transformation, while the background gauge
field $\overline{V}_\mu$ transforms inhomogeneously:
\begin{eqnarray}
&& \overline{\xi}_{\rm R,L} \rightarrow h(x) \overline{\xi}_{\rm R,L} 
  g_{\rm R,L}^\dag \ ,
\nonumber\\
&& \overline{V}_\mu \rightarrow h(x) \overline{V}_\mu h^\dag(x)
+ i h(x) \cdot \partial_\mu h^\dag(x) \ ,
\nonumber\\
&& \widehat{\varphi}_\pi \rightarrow 
  h(x) \widehat{\varphi}_\pi h^\dag(x) \ ,
\nonumber\\
&& \widehat{\varphi}_\sigma \rightarrow 
  h(x) \widehat{\varphi}_\sigma h^\dag(x) \ ,
\nonumber\\
&& \widehat{v}_\mu \rightarrow h(x) \widehat{v}_\mu h^\dag(x) \ .
\end{eqnarray}
Thus, the expansion of the Lagrangian in terms of the quantum field
does not violate the HLS of the background field
$\overline{V}_\mu$~\cite{Tanabashi}.

We adopt 
the background gauge fixing
in 't~Hooft--Feynman gauge,
\begin{eqnarray}
{\cal L}_{\rm GF} =
- \mbox{tr}\,
\biggl[
\left(
  \overline{D}^\mu \hat{v}_\mu + g F_\sigma^2 \widehat{\varphi}_\sigma 
\right)^2
\biggr]
\ ,
\label{gauge fixing}
\end{eqnarray}
where $\overline{D}_\mu$ is the covariant derivative on the background
field:
\begin{equation}
\overline{D}^\mu \hat{v}_\nu = \partial^\mu \hat{v}_\nu
- i \left[ \overline{V}^\mu , \hat{v}_\nu \right] \ .
\end{equation}
The Faddeev-Popov (FP) ghost term associated with the gauge 
fixing~(\ref{gauge fixing}) is
\begin{equation}
{\cal L}_{\rm FP} = 
2 i \, \mbox{tr} \, 
\left[
  \overline{C}
  \left( 
    \overline{D}^\mu \overline{D}_\mu + (g F_\sigma)^2 
  \right)
  C
\right]
+ \cdots
\ ,
\end{equation}
where the ellipsis stands for interaction terms of the dynamical
fields
$\widehat{\varphi}_\pi$, $\widehat{\varphi}_\sigma$, and
$\widehat{v}_\mu$ and the FP ghosts.

Now, the complete ${\cal O}(p^2)$ Lagrangian
${\cal L}_{(2)} + {\cal L}_{\rm GF} + {\cal L}_{\rm FP}$ is
expanded in terms of the quantum fields $\widehat{\varphi}_\pi$, 
$\widehat{\varphi}_\sigma$, $\widehat{v}$ and $C$.
The terms which do not include the quantum fields are nothing but the
original ${\cal O}(p^2)$ Lagrangian with the fields replaced by the
corresponding background fields.
The terms which are of first order in the quantum fields 
lead to the equations of motions for the background fields:
\begin{eqnarray}
&&
\overline{D}_\mu \overline{\cal A}^\mu =
- i \left( a- 1 \right)
\left[
  \overline{\cal V}_\mu - \overline{V}_\mu \,,\,
  \overline{\cal A}^\mu
\right]
\nonumber\\
&& \quad
+ \frac{i}{4} \frac{F_\chi^2}{F_\pi^2} 
\left(
  \overline{\chi} - \overline{\chi}^\dag
  - \frac{1}{N_f} \, \mbox{tr}
  \left[ \overline{\chi} - \overline{\chi}^\dag \right]
\right)
+ {\cal O}(p^4)
\ ,
\label{EOM Npi B}
\\
&&
\overline{D}_\mu 
\left( \overline{\cal V}^\mu - \overline{V}^\mu \right)
= {\cal O}(p^4)
\ ,
\label{EOM Nsig B}
\\
&&
\overline{D}_\nu \overline{V}^{\nu\mu}
=
g^2 F_\sigma^2 
\left( \overline{\cal V}^\mu - \overline{V}^\mu \right)
+ {\cal O}(p^4)
\ ,
\label{EOM Nvec B}
\end{eqnarray}
which correspond to Eqs.~(\ref{EOM Npi2}), (\ref{EOM Nsig2}) and
(\ref{EOM Nvec2}), respectively.

To write down the terms which are of quadratic order in the quantum
fields in a compact and unified way,
let us define the following ``connections'':
\begin{eqnarray}
\Gamma^{(\pi\pi)}_{\mu,ab} &\equiv&
i \, \mbox{tr} \Biggl[
  \biggl(
    \left( 2 - a \right) \overline{\cal V}_\mu
    + a \overline{V}_\mu 
  \biggr)
  \left[ T_a \,,\, T_b \right]
\Biggr]
\ ,
\label{Xpp}
\\
\Gamma^{(\sigma\sigma)}_{\mu,ab} &\equiv&
i \, \mbox{tr} \Biggl[
  \biggl(
    \overline{\cal V}_\mu + \overline{V}_\mu 
  \biggr)
  \left[ T_a \,,\,T_b \right]
\Biggr]
\ ,
\label{Xss}
\\
\Gamma^{(\pi\sigma)}_{\mu,ab} &\equiv&
- i \sqrt{a} \, \mbox{tr} \biggl[
  \overline{\cal A}_\mu
  \left[ T_a \,,\, T_b \right]
\biggr]
\ ,
\label{Xps}
\\
\Gamma^{(\sigma\pi)}_{\mu,ab} &\equiv&
- i \sqrt{a} \, \mbox{tr} \biggl[
  \overline{\cal A}_\mu
  \left[ T_a \,,\, T_b \right]
\biggr]
\ ,
\label{Xsp}
\\
\Gamma^{(V_\alpha V_\beta)}_{\mu,ab} &\equiv&
- 2 i \, \mbox{tr} \biggl[
  \overline{V}_\mu \left[ T_a \,,\,T_b \right]
\biggr] \, g^{\alpha\beta}
\ .
\label{Xvv}
\end{eqnarray}
Here one might doubt the minus sign in
front of $\Gamma_\mu^{(V_\alpha V_\beta)}$ compared with
$\Gamma_\mu^{(SS)}$
($S = \pi,\sigma$).  However, since $g^{\alpha\beta} = -
\delta_{\alpha\beta}$ for $\alpha = 1$, 2, 3, the minus sign is the
correct one.
Correspondingly, we should use an unconventional metric 
$- g_{\alpha\beta}$ to change the upper indices to the lower ones:
\begin{eqnarray}
{\Gamma_\mu}^{V_\beta)}_{(V_\alpha,ab} &\equiv&
\sum_{\alpha'} \left( - g_{\alpha \alpha'} \right)
{\Gamma_\mu}^{(V_{\alpha'} V_{\beta})}_{ab}
\ .
\label{Xv lu}
\end{eqnarray}
Further we define the following quantities corresponding to the
``mass'' part:
\begin{eqnarray}
&&
\Sigma^{(\pi\pi)}_{ab} \equiv
-\frac{4-3a}{2}
\, \mbox{tr} \biggl[
  \left[ \overline{\cal A}^\mu \,,\, T_a \right]
  \left[ \overline{\cal A}_\mu \,,\, T_b \right]
\biggr]
\nonumber\\
&&\ 
{}- \frac{a^2}{2} \mbox{tr} \biggl[
  \left[ \overline{\cal V}^\mu - \overline{V}^\mu \,,\, T_a \right]
  \left[ \overline{\cal V}_\mu - \overline{V}_\mu \,,\, T_b
  \right]
\biggr]
\nonumber\\
&&\ 
{} + \frac{F_\chi^2}{2F_\pi^2} \, \mbox{tr}
\biggl[
  \left( 
    \overline{\chi} + \overline{\chi}^\dag - 2 \widehat{\cal M}_\pi
  \right)
  \left\{ T_a \,,\, T_b \right\}
\biggr]
\ ,
\label{Ypp}
\\
&&
\Sigma^{(\sigma\sigma)}_{ab} \equiv
- \frac{1}{2} \, \mbox{tr} \biggl[
  \left[ 
    \overline{\cal V}^\mu - \overline{V}^\mu \,,\, T_a
  \right]
  \left[ 
    \overline{\cal V}_\mu - \overline{V}_\mu \,,\, T_b
  \right]
\biggr]
\nonumber\\
&&\ 
{}-
\frac{a}{2} 
\, \mbox{tr} \biggl[
  \left[ 
    \overline{\cal A}^\mu \,,\, T_a
  \right]
  \left[ 
    \overline{\cal A}_\mu \,,\, T_b
  \right]
\biggr]
\ ,
\label{Yss}
\\
&&
\Sigma^{(\pi\sigma)}_{ab} \equiv
i \sqrt{a}\, \mbox{tr} \biggl[
  \overline{D}^\mu \overline{\cal A}_\mu
  \left[ T_a \,,\, T_b \right]
\biggr]
\nonumber\\
&&\ 
{}+ \frac{1}{2} \sqrt{a}\, \mbox{tr} \biggl[
  \left[ 
    \overline{\cal A}_\mu \,,\, T_a
  \right]
  \left[ 
    \overline{\cal V}^\mu - \overline{V}^\mu \,,\, T_b
  \right]
\biggr]
\nonumber\\
&&\ 
{}+ \left( 1 - \frac{a}{2} \right) 
\sqrt{a}\, \mbox{tr}
\biggl[
  \left[ 
    \overline{\cal V}_\mu - \overline{V}_\mu \,,\, T_a
  \right]
  \left[ 
    \overline{\cal A}^\mu \,,\, T_b
  \right]
\biggr]
\ ,
\\
&&
\Sigma^{(\sigma\pi)}_{ab} \equiv \Sigma^{(\pi\sigma)}_{ba} 
\ ,
\\
&&
\Sigma^{(V_\alpha V_\beta)}_{ab} \equiv
- 4 i \mbox{tr} 
\biggl[
  \overline{V}^{\alpha\beta} \left[ T_a \,,\, T_b \right]
\biggr]
\ ,
\label{Yvv}
\\
&&
\Sigma^{(\pi V_\beta)}_{ab} \equiv
2 i a g F_\pi \, \mbox{tr} \biggl[
  \overline{\cal A}^\beta
  \left[ 
     T_a \,,\, T_b
  \right]
\biggr]
\ ,
\label{Ypv}
\\
&&
\Sigma^{(V_\alpha \pi)}_{ab} \equiv
- 2 i a g F_\pi \, \mbox{tr} \biggl[
  \overline{\cal A}^\alpha
  \left[ 
     T_a \,,\, T_b
  \right]
\biggr]
\ ,
\label{Yvp}
\\
&&
\Sigma^{(\sigma V_\beta)}_{ab} \equiv
2 i g F_\sigma \mbox{tr} \biggl[
  \left( \overline{\cal V}^\beta - \overline{V}^\beta \right)
  \left[ 
     T_a \,,\, T_b
  \right]
\biggr]
\ ,
\label{Ysv}
\\
&&
\Sigma^{(V_\alpha \sigma)}_{ab} \equiv
- 2 i g F_\sigma \mbox{tr} \biggl[
  \left( \overline{\cal V}^\alpha - \overline{V}^\alpha \right)
  \left[ 
     T_a \,,\, T_b
  \right]
\biggr]
\ ,
\label{Yvs}
\end{eqnarray}
where
\begin{equation}
{\cal M}_\pi \equiv 2 B {\cal M}
\ ,
\end{equation}
with the quark mass matrix ${\cal M}$ being defined in 
Eq.~(\ref{quark mass matrix}).
Here by using the equation of motion in Eq.~(\ref{EOM Npi B}),
$\Sigma^{(\pi\sigma)}_{ab}$ is rewritten as
\begin{eqnarray}
&&
\Sigma^{(\pi\sigma)}_{ab}
=
\sqrt{a} (1-a) \, \mbox{tr}
\biggl[
  \left[ 
    \overline{\cal A}_\mu \,,\,
    \overline{\cal V}^\mu - \overline{V}^\mu
  \right]
  \left[ T_a \,,\, T_b \right]
\biggr]
\nonumber\\
&&\ 
- i\, \frac{\sqrt{a}}{4} \, \frac{F_\chi^2}{F_\pi^2} 
\, \mbox{tr}
\biggl[
  \left( \overline{\chi} - \overline{\chi}^\dag \right)
  \left[ T_a \,,\, T_b \right]
\biggr]
\nonumber\\
&&\ 
{}+ \frac{1}{2} \sqrt{a}\, \mbox{tr} \biggl[
  \left[ 
    \overline{\cal A}_\mu \,,\, T_a
  \right]
  \left[ 
    \overline{\cal V}^\mu - \overline{V}^\mu \,,\, T_b
  \right]
\biggr]
\nonumber\\
&&\ 
{}+ \left( 1 - \frac{a}{2} \right) 
\sqrt{a}\, \mbox{tr}
\biggl[
  \left[ 
    \overline{\cal V}_\mu - \overline{V}_\mu \,,\, T_a
  \right]
  \left[ 
    \overline{\cal A}^\mu \,,\, T_b
  \right]
\biggr]
\ .
\label{Yps:eom}
\end{eqnarray}

To achieve more unified treatment let us introduce the following
quantum fields:
\begin{equation}
\widehat{\Phi}_A \equiv
\left( \widehat{\pi}^a \,,\,
\widehat{\sigma}^a \,,\,
\widehat{v}_{\alpha}^a 
\right)
\equiv
\left( F_\pi \widehat{\varphi}_\pi^a \,,\,
F_\sigma \widehat{\varphi}_\sigma^a \,,\,
\widehat{v}_{\alpha}^a 
\right) \ ,
\end{equation}
where the lower and upper indices of
$\widehat{\Phi}$ should be distinguished as in Eq.~(\ref{Xv lu}).
Thus the metric acting on the indices of $\widehat{\Phi}$ is defined
by
\begin{eqnarray}
\eta^{AB} &\equiv&
\left( \begin{array}{ccc}
\delta_{ab} & & \\
& \delta_{ab} & \\
& & - g^{\alpha\beta} \delta_{ab}
\end{array} \right)
\ ,
\nonumber\\
\eta^{A}_{B} &\equiv&
\left( \begin{array}{ccc}
\delta_{ab} & & \\
& \delta_{ab} & \\
& & g^{\alpha}_{\beta} \delta_{ab}
\end{array} \right)
\ ,
\nonumber\\
\eta_{AB} &\equiv&
\left( \begin{array}{ccc}
\delta_{ab} & & \\
& \delta_{ab} & \\
& & - g_{\alpha\beta} \delta_{ab}
\end{array} \right)
\ .
\end{eqnarray}
The tree mass matrix is defined by
\begin{equation}
{\cal M}^{AB} \equiv
\left( \begin{array}{ccc}
\bar{M}_{\pi,a} \delta_{ab} & & \\
& \bar{M}_V^2 \delta_{ab} & \\
& & - g^{\alpha\beta} \bar{M}_V^2 \delta_{ab}
\end{array} \right)
\ ,
\end{equation}
where $\bar{M}_V^2 \equiv g^2 F_\sigma^2$, and 
the pseudoscalar meson mass $\bar{M}_{\pi,a}$ is defined by
\begin{equation}
\bar{M}_{\pi,a}^2 \delta_{ab} \equiv
\frac{F_\chi^2}{F_\pi^2} \, \mbox{tr}
\left[
  \widehat{\cal M}_\pi 
  \left\{ T_a \,,\, T_b \right\}
\right]
\ .
\end{equation}
Here the generator $T_a$ is defined in such a way
that the above masses are diagonalized when we introduce the explicit
chiral symmetry breaking due to the current quark masses.
It should be noticed that we work in the chiral limit in this paper,
so that we take 
\begin{equation}
{\cal M}_\pi = 0 \quad \mbox{or} \quad
\bar{M}_{\pi,a}  = 0 \ .
\end{equation}

Let us further define
\begin{eqnarray}
\left( \widetilde{\Gamma}_\mu \right)^{AB} 
&\equiv&
\left( \begin{array}{ccc}
\Gamma_{\mu,ab}^{(\pi\pi)} & \Gamma_{\mu,ab}^{(\pi\sigma)} & 0 \\
\Gamma_{\mu,ab}^{(\sigma\pi)} & \Gamma_{\mu,ab}^{(\sigma\sigma)} & 0 \\
0 & 0 & \Gamma^{(V_\alpha V_\beta)}_{\mu,ab}
\end{array} \right)
\ ,
\\
\widetilde{\Sigma}^{AB} 
&\equiv&
\left( \begin{array}{ccc}
\Sigma_{ab}^{(\pi\pi)} & \Sigma_{ab}^{(\pi\sigma)} & 
   \Sigma_{ab}^{(\pi V_\beta)} \\
\Sigma_{ab}^{(\sigma\pi)} & \Sigma_{ab}^{(\sigma\sigma)} & 
   \Sigma_{ab}^{(\sigma V_\beta)} \\
\Sigma_{ab}^{(V_\alpha\pi)} & \Sigma_{ab}^{(V_\alpha\sigma)} & 
   \Sigma_{ab}^{(V_\alpha V_\beta)}
\end{array} \right)
\ ,
\end{eqnarray}
and
\begin{equation}
\left( \widetilde{D}_\mu \right)^{AB} \equiv
\eta^{AB} \partial_\mu + 
\left( \widetilde{\Gamma}_\mu \right)^{AB} \ .
\end{equation}
It is convenient to consider the FP ghost contribution separately.
For the FP ghost part we define similar quantities:
\begin{eqnarray}
\Gamma^{(CC)}_{\mu,ab} &\equiv&
2 i \, \mbox{tr} \biggl[
  \overline{V}_\mu \left[ T_a \,,\,T_b \right]
\biggr]
\ ,
\label{XCC}
\\
\left( \widetilde{D}_\mu\right)_{ab}^{(CC)} &\equiv&
\delta_{ab} \partial_\mu + \Gamma^{(CC)}_{\mu,ab}
\ ,
\label{def:DCC}
\\
\widetilde{\cal M}^{(CC)}_{ab} &\equiv&
\delta_{ab} \bar{M}_V^2
\ .
\end{eqnarray}
By using the above quantities the terms quadratic in terms of the
quantum fields in the total Lagrangian
are rewritten as
\begin{eqnarray}
&&
\int \! d^4x\, 
\left[ 
  {\cal L}_{(2)}^{(2)} + {\cal L}_{\rm GF} + {\cal L}_{\rm FP}
\right]
=
\nonumber\\
&&\ 
- \frac{1}{2} \sum_{A,B} \int \! d^4x \,
\widehat{\Phi}_A
\biggl[
  \left( \widetilde{D}_\mu \cdot \widetilde{D}^\mu \right)^{AB}
  + \widetilde{\cal M}^{AB}
  + \widetilde{\Sigma}^{AB} 
\biggr]
\widehat{\Phi}_B
\nonumber\\
&& \ 
{}+ i \sum_{a,b} \int \! d^4x \, \overline{C}^a
\biggl[
  \left( 
     \widetilde{D}_{\mu} \cdot \widetilde{D}^{\mu} 
  \right)_{ab}^{(CC)}
  + \widetilde{\cal M}^{(CC)}_{ab}
\biggr]
C^b
\ ,
\end{eqnarray}
where
\begin{eqnarray}
&&
\left( \widetilde{D}_\mu \cdot \widetilde{D}^\mu \right)^{AB}
\equiv
\sum_{A'} \left( \widetilde{D}_\mu \right)^{AA'}
\left( \widetilde{D}^\mu \right)^{B}_{A'}
\ ,
\\
&&
\left( 
   \widetilde{D}_{\mu} \cdot \widetilde{D}^{\mu} 
\right)_{ab}^{(CC)}
\equiv
\sum_{c}
\left( \widetilde{D}_\mu \right)^{(CC)}_{ac}
\left( \widetilde{D}_\mu \right)^{(CC)}_{cb}
\ .
\label{Lag:BGF}
\end{eqnarray}

\section{Renormalization Group Equations}
\label{sec:RGE}

In this appendix, we show the detailed derivation of
the RGE's for
$F_\pi$, $F_\sigma$ (and $a\equiv F_\sigma^2/F_\pi^2$),
$g$, $z_1$, $z_2$ and $z_3$ for the reader's convenience.
These RGE's are derived by calculating the divergent
corrections at one loop to the
two-point functions of the background fields, $\overline{\cal A}_\mu$,
$\overline{\cal V}_\mu$ and $\overline{V}_\mu$.
Note that the RGE's for 
$F_\pi$, $a\equiv F_\sigma^2/F_\pi^2$ and $g$ without quadratic
divergences were obtained in Ref.~\cite{HY}.
Note also that
the RGE's for $F_\pi$ and $a$ 
with quadratic divergences were derived in
Ref.~\cite{HY:letter},
and the RGE's for $z_1$, $z_2$ and $z_3$ were in
Ref.~\cite{Tanabashi}.

In the present analysis it is important to include
{\it quadratic divergences} to obtain RGE's
in the Wilsonian sense.
Since a naive momentum cutoff violates chiral symmetry,
we need a careful treatment of the quadratic divergences.
Thus we adopt dimensional regularization
and identify quadratic divergences with the presence of poles
of ultraviolet origin at $n=2$~\cite{Veltman}.
This can be done by the following replacement in the Feynman
integrals:
\begin{eqnarray}
&&
\int \frac{d^n k}{i (2\pi)^n} \frac{1}{-k^2} \rightarrow 
\frac{\Lambda^2} {(4\pi)^2} \ ,
\nonumber\\
&&
\int \frac{d^n k}{i (2\pi)^n} 
\frac{k_\mu k_\nu}{\left[-k^2\right]^2} \rightarrow 
- \frac{\Lambda^2} {2(4\pi)^2} g_{\mu\nu} \ .
\label{replace}
\end{eqnarray}
On the other hand, 
the logarithmic divergence is identified with the pole at 
$n=4$.
The same result as that after the replacements Eq.~(\ref{replace})
can also be obtained in the heat kernel expansion with the proper time
regularization in which the physical interpretation 
of the quadratic divergence is more explicit with $\Lambda$ 
having the same meaning as the naive cutoff.

\begin{figure}[bthp]
\begin{center}
\epsfxsize = 8cm
\  \epsfbox{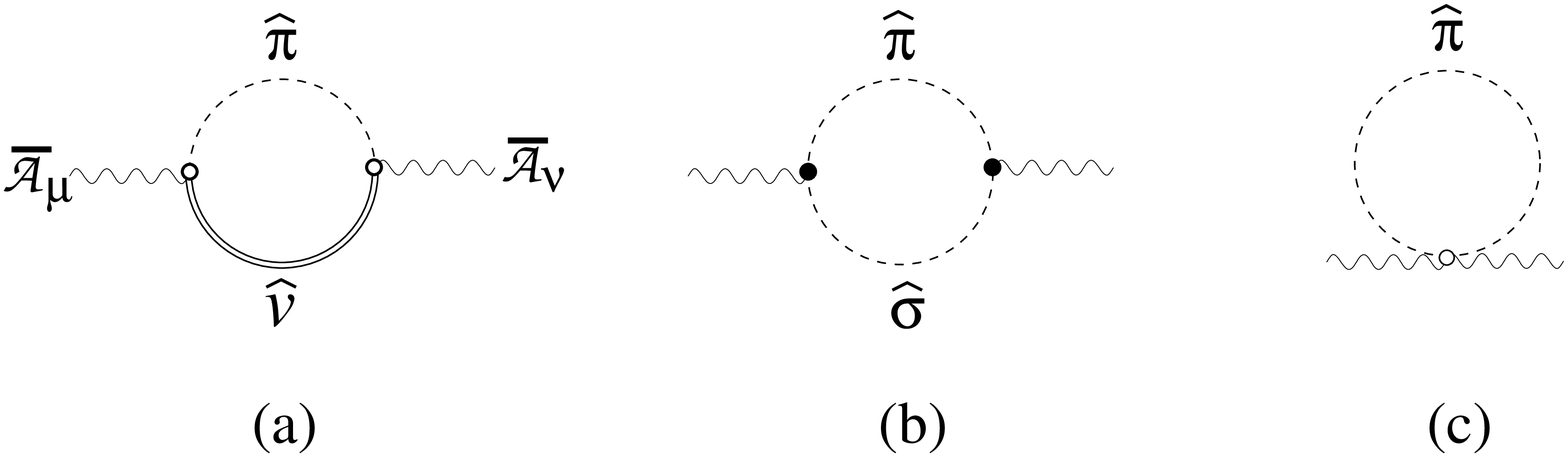}
\end{center}
\caption[$\overline{\cal A}_\mu$-$\overline{\cal A}_\nu$ Two-Point 
function]{%
One-loop corrections to the two-point function
$\overline{\cal A}_\mu$-$\overline{\cal A}_\nu$.
The vertex with a dot ($\bullet$) implies the derivatives
acting on the quantum fields, while that with a circle ($\circ$)
implies no derivatives are included:
The vertices in (a) are from $\Sigma^{(\pi V_\beta)}_{ab}$
and $\Sigma^{(V_\alpha \pi)}_{ab}$ in Eqs.~(\ref{Ypv}) and
(\ref{Yvp});
the vertices in (b) are from $\Gamma^{(\pi\sigma)}_{\mu,ab}$ and
$\Gamma^{(\sigma\pi)}_{\mu,ab}$ in Eqs.~(\ref{Xps}) and (\ref{Xsp})
together with the derivatives acting on the quantum fields;
the vertex in (c) is from the first term of $\Sigma^{(\pi\pi)}_{ab}$
in Eq.~(\ref{Ypp}) 
and $\sum_c \Gamma^{(\pi\sigma)}_{\mu,ac} 
\Gamma^{\mu,(\sigma\pi)}_{cb}$~\cite{foot:Yss}.
}\label{fig:aa}
\end{figure}
Let us start from the one-loop corrections to the
two-point function
$\overline{\cal A}_\mu$-$\overline{\cal A}_\nu$.
The relevant diagrams are shown in Fig.~\ref{fig:aa}.
The divergent contributions of these diagrams are evaluated as
\begin{eqnarray}
&&
\left.
\Pi_{\overline{\cal A}\overline{\cal A}}^{{\rm(a)}\mu\nu}(p)
\right\vert_{\rm div}
= 
g^{\mu\nu}\,\frac{N_f}{2(4\pi)^2}
\left[
  - 2 a \bar{M}_V^2\, \ln \Lambda^2
\right]
\ ,
\nonumber\\
&&
\left.
\Pi_{\overline{\cal A}\overline{\cal A}}^{{\rm(b)}\mu\nu}(p)
\right\vert_{\rm div}
=
g^{\mu\nu}\, \frac{N_f}{2(4\pi)^2}
\left[
  - a \Lambda^2 + \frac{1}{2} a \bar{M}_V^2\, \ln \Lambda^2
\right]
\nonumber\\
&&\quad
{}- \left( g^{\mu\nu} p^2 - p^\mu p^\nu \right)
\, \frac{N_f}{2(4\pi)^2} \frac{a}{6} \ln \Lambda^2
\ ,
\nonumber\\
&&
\left.
\Pi_{\overline{\cal A}\overline{\cal A}}^{{\rm(c)}\mu\nu}(p)
\right\vert_{\rm div}
=
g^{\mu\nu}\, \frac{N_f}{2(4\pi)^2}
\left[
  2 (a-1) \Lambda^2
\right]
\ .
\label{div:aa}
\end{eqnarray}
The divergences in Eq.~(\ref{div:aa})
are renormalized by the bare parameters in the
Lagrangian.
The tree level contribution with the bare parameters
is given by
\begin{equation}
\Pi_{\overline{\cal A}\overline{\cal A}}^{{\rm(tree)}\mu\nu}(p^2)
= F_{\pi,{\rm bare}}^2 \, g^{\mu\nu} 
+ 2 z_{2,{\rm bare}} \left( p^2 g^{\mu\nu} - p^\mu p^\nu \right) \ .
\end{equation}
Thus the renormalization is done by requiring 
that the followings be finite:
\begin{eqnarray}
&&
F_{\pi,{\rm bare}}^2 
- \frac{N_f}{4(4\pi)^2} \left[
  2 (2-a) \Lambda^2 + 3 a^2 g^2 F_\pi^2 \ln \Lambda^2
\right]
\nonumber\\
&&\qquad\qquad
= \mbox{(finite)} \ ,
\nonumber\\
&&
z_{2,{\rm bare}} 
- 
\frac{N_f}{2(4\pi)^2} \, \frac{a}{12} \ln \Lambda^2
= \mbox{(finite)}
\ .
\end{eqnarray}
The above renormalizations lead to the following RGE's
for $F_\pi$ [the first equation in Eqs.~(\ref{RGE for g2})]
and $z_2$ [the second equation in Eqs.~(\ref{RGE z})]:
\begin{eqnarray}
&&
\mu \frac{d F_\pi^2}{d\mu} =
\frac{N_f}{2(4\pi)^2}
\left[ 3 a^2 g^2 F_\pi^2 + 2 (2-a) \mu^2 \right] \ ,
\label{app:RGE:Fp}
\\
&&
\mu \frac{d z_2}{d \mu} =
\frac{N_f}{(4\pi)^2} \frac{a}{12} \ ,
\label{app:RGE:z2}
\end{eqnarray}
where $\mu$ is the renormalization scale.

\begin{figure}[htbp]
\begin{center}
\epsfxsize = 8cm
\  \epsfbox{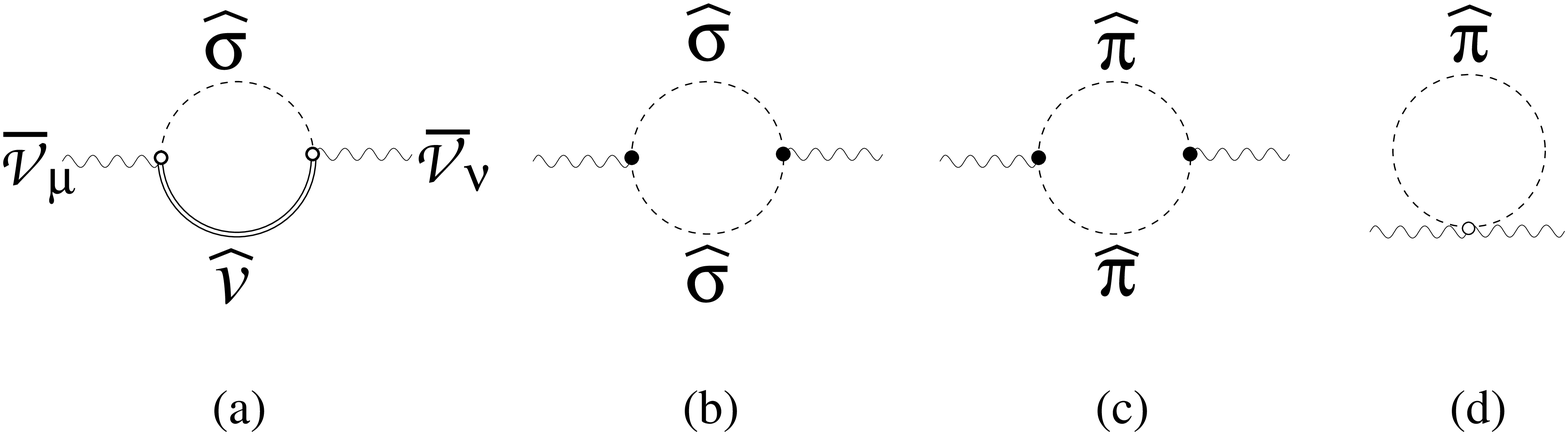}
\end{center}
\caption[$\overline{\cal V}_\mu$-$\overline{\cal V}_\nu$ Two-Point 
function]{%
One-loop corrections to the two-point function
$\overline{\cal V}_\mu$-$\overline{\cal V}_\nu$.
The vertices in (a) are from $\Sigma^{(\sigma V_\beta)}_{ab}$ and
$\Sigma^{(V_\alpha \sigma)}_{ab}$ in Eqs.~(\ref{Ysv}) and (\ref{Yvs});
the vertices in (b) are from $\Gamma^{(\sigma\sigma)}_{\mu,ab}$ in 
Eq.~(\ref{Xss}) together with derivatives acting on the quantum
fields;
the vertices in (c) are from $\Gamma^{(\pi\pi)}_{\mu,ab}$ in 
Eq.~(\ref{Xpp}) together with derivatives acting on the quantum
fields;
the vertex in (d) is from the second term of $\Sigma^{(\pi\pi)}_{ab}$
in Eq.~(\ref{Ypp}) and $\sum_c \Gamma^{(\pi\pi)}_{\mu,ac} 
\Gamma^{\mu,(\pi\pi)}_{cb}$.
}\label{fig:vv}
\end{figure}
Next we calculate one-loop corrections to the
two-point function
$\overline{\cal V}_\mu$-$\overline{\cal V}_\nu$.
The relevant diagrams are shown in Fig.~\ref{fig:vv}.
The divergent contributions are evaluated as
\begin{eqnarray}
&&
\left.
\Pi_{\overline{\cal V}\overline{\cal V}}^{{\rm(a)}\mu\nu}(p)
\right\vert_{\rm div}
=
g^{\mu\nu}\,\frac{N_f}{2(4\pi)^2}
\left[
  - 2 a \bar{M}_V^2\, \ln \Lambda^2
\right]
\ ,
\nonumber\\
&&
\left.
\Pi_{\overline{\cal V}\overline{\cal V}}^{{\rm(b)}\mu\nu}(p)
\right\vert_{\rm div}
=
g^{\mu\nu}\,\frac{N_f}{2(4\pi)^2}
\left[
  - \frac{1}{2} \Lambda^2
  + \frac{1}{2} \bar{M}_V^2\, \ln \Lambda^2
\right]
\nonumber\\
&&\quad
{}- \left( g^{\mu\nu} p^2 - p^\mu p^\nu \right)
\, \frac{N_f}{2(4\pi)^2} \frac{1}{12} \ln \Lambda^2
\ ,
\nonumber\\
&&
\left.
\Pi_{\overline{\cal V}\overline{\cal V}}^{{\rm(c)}\mu\nu}(p)
\right\vert_{\rm div}
=
g^{\mu\nu}\,\frac{N_f}{2(4\pi)^2}
\left[
  - \frac{(2-a)^2}{2} \Lambda^2
\right]
\nonumber\\
&&\quad
{}- \left( g^{\mu\nu} p^2 - p^\mu p^\nu \right)
\, \frac{N_f}{2(4\pi)^2} \frac{(2-a)^2}{12} \ln \Lambda^2
\ ,
\nonumber\\
&&
\left.
\Pi_{\overline{\cal V}\overline{\cal V}}^{{\rm(d)}\mu\nu}(p)
\right\vert_{\rm div}
=
g^{\mu\nu}\,\frac{N_f}{2(4\pi)^2}
\left[
  - 2 (a-1) \Lambda^2
\right]
\ .
\end{eqnarray}
Similarly to the 
$\overline{\cal A}_\mu$-$\overline{\cal A}_\nu$ two-point function,
we require that the following quantities be finite:
\begin{eqnarray}
&&
F_{\sigma,{\rm bare}}^2 
- \frac{N_f}{4(4\pi)^2} \left[
  (1+a^2) \Lambda^2 + 3 a g^2 F_\pi^2 \ln \Lambda^2
\right]
\nonumber\\
&&\qquad\qquad
= \mbox{(finite)} \ ,
\nonumber
\\
&&
z_{1,{\rm bare}} 
- 
\frac{N_f}{2(4\pi)^2} \, \frac{5-4a+a^2}{12} \ln \Lambda^2
= \mbox{(finite)}
\ .
\label{Fs:z1:renorm}
\end{eqnarray}
The above renormalizations lead to the following RGE's
for $F_\sigma$ and $z_1$ [the first equation in Eqs.~(\ref{RGE z})]:
\begin{eqnarray}
&&
\mu \frac{d F_\sigma^2}{d\mu} =
\frac{N_f}{2(4\pi)^2}
\left[ 3 a g^2 F_\pi^2 + (1+a^2) \mu^2 \right] \ ,
\label{app:RGE:Fs}
\\
&&
\mu \frac{d z_1}{d \mu} =
\frac{N_f}{(4\pi)^2} \frac{5-4a+a^2}{24} \ .
\label{app:RGE:z1}
\end{eqnarray}
The RGE for $a\equiv F_\sigma^2/F_\pi^2$
[the second equation in Eqs.~(\ref{RGE for g2})]
is derived from the RGE's for $F_\sigma$ and $F_\pi$
given in Eqs.~(\ref{app:RGE:Fp}) and (\ref{app:RGE:Fs}):
\begin{equation}
\mu \frac{d a}{d\mu} = - 
C
(a-1)
\left[ 3 a (a+1) g^2 - (3a-1) \frac{\mu^2}{F_\pi^2} \right] \ ,
\label{app:RGE:a}
\end{equation}
where $C = N_f/[2(4\pi)^2]$.

\begin{figure}[htbp]
\begin{center}
\epsfxsize = 8cm
\  \epsfbox{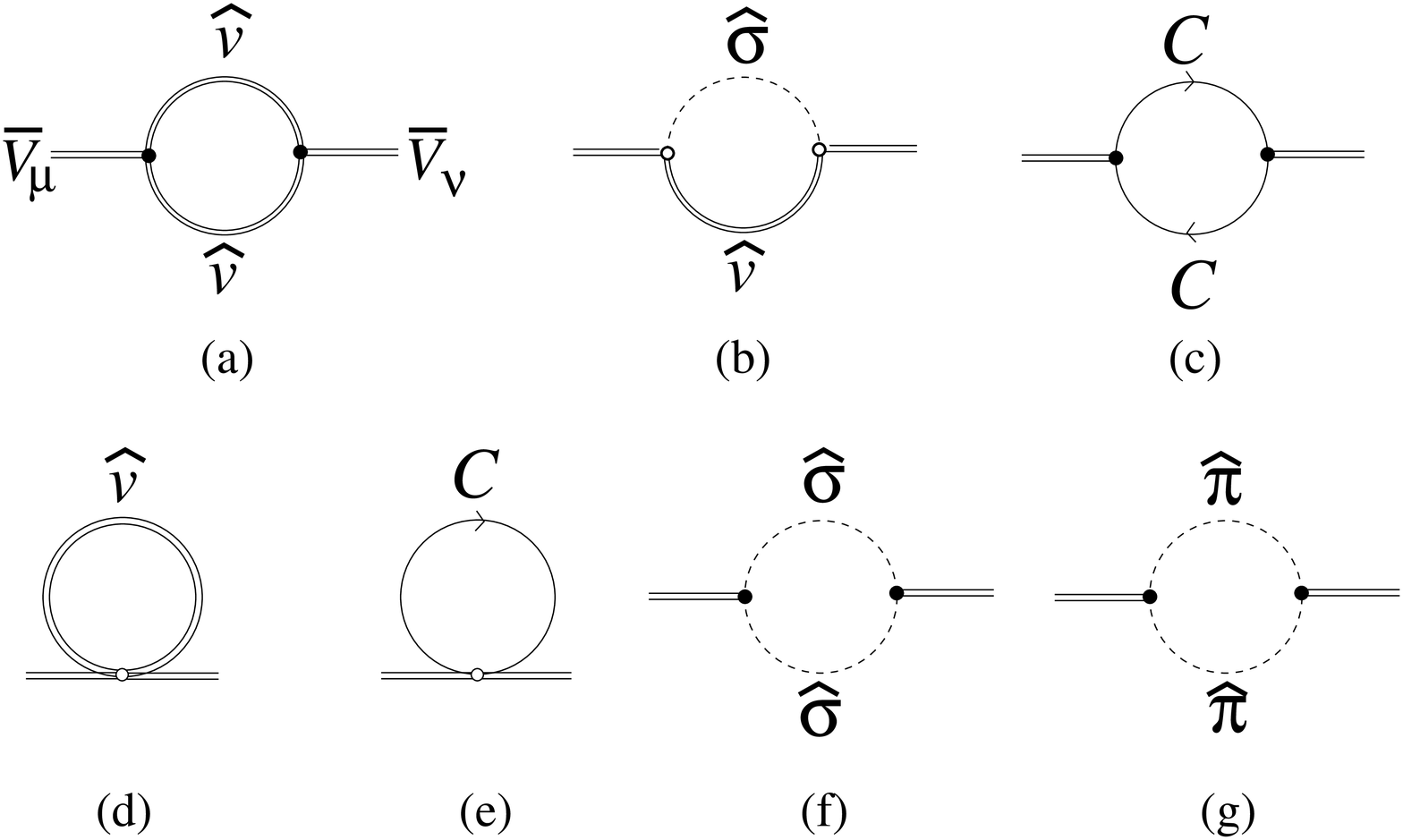}
\end{center}
\caption[$\overline{V}_\mu$-$\overline{V}_\nu$ Two-Point 
function]{%
One-loop corrections to the two-point function
$\overline{V}_\mu$-$\overline{V}_\nu$.
The vertices in (a) are from $\Sigma^{(V_\alpha V_\beta)}_{ab}$ in 
Eq.~(\ref{Yvv}) and $\Gamma^{(V_\alpha V_\beta)}_{\mu,ab}$ in
Eq.~(\ref{Xvv}) together with derivatives acting on the quantum
fields;
the vertices in (b) are from $\Sigma^{(\sigma V_\beta)}_{ab}$ and
$\Sigma^{(V_\alpha \sigma)}_{ab}$ in Eqs.~(\ref{Ysv}) and (\ref{Yvs});
the vertices in (c) are from $\Gamma^{(CC)}_{\mu,ab}$ in
Eq.~(\ref{XCC}) 
together with derivatives acting on the quantum fields;
the vertex in (d) is from $\Sigma^{(V_\alpha V_\beta)}_{ab}$
and $\sum_{c,\gamma} \Gamma^{(V_\alpha }_{\mu,V_\gamma),ac}
{\Gamma^{\mu,(V_\gamma V_\beta)}_{cb}}$;
the vertex in (e) is from $\sum_c \Gamma^{(CC)}_{\mu,ac} 
\Gamma^{\mu,(CC)}_{cb}$;
the vertices in (f) are from $\Gamma^{(\sigma\sigma)}_{\mu,ab}$ in
Eq.~(\ref{Xss}) together with derivatives acting on the quantum
fields;
the vertices in (g) are from $\Gamma^{(\pi\pi)}_{\mu,ab}$ in
Eq.~(\ref{Xpp})
together with derivatives acting on the quantum fields.
}\label{fig:rr}
\end{figure}

Now, we calculate the one-loop correction to the
two-point function
$\overline{V}_\mu$-$\overline{V}_\nu$.
The relevant diagrams are shown in Fig.~\ref{fig:rr}.
These are evaluated as
\begin{eqnarray}
&&
\left.
\Pi_{\overline{V}\overline{V}}^{{\rm(a)}\mu\nu}(p)
\right\vert_{\rm div}
=
g^{\mu\nu}\,\frac{N_f}{2(4\pi)^2}
\left[
  - 4 \Lambda^2
  + 8 \bar{M}_V^2\, \ln \Lambda^2
\right]
\nonumber\\
&&\quad
{}+ \left( g^{\mu\nu} p^2 - p^\mu p^\nu \right)
\, \frac{N_f}{2(4\pi)^2} \frac{20}{3} \ln \Lambda^2
\ ,
\nonumber\\
&&
\left.
\Pi_{\overline{V}\overline{V}}^{{\rm(b)}\mu\nu}(p)
\right\vert_{\rm div}
=
g^{\mu\nu}\,\frac{N_f}{2(4\pi)^2}
\left[
  - 2 \bar{M}_V^2\, \ln \Lambda^2
\right]
\ ,
\nonumber\\
&&
\left.
\Pi_{\overline{V}\overline{V}}^{{\rm(c)}\mu\nu}(p)
\right\vert_{\rm div}
=
g^{\mu\nu}\,\frac{N_f}{2(4\pi)^2}
\left[
  4 \Lambda^2
  - 4 \bar{M}_V^2\, \ln \Lambda^2
\right]
\nonumber\\
&&\quad
{}+ \left( g^{\mu\nu} p^2 - p^\mu p^\nu \right)
\, \frac{N_f}{2(4\pi)^2} \frac{2}{3} \ln \Lambda^2
\ ,
\nonumber\\
&&
\left.
\Pi_{\overline{V}\overline{V}}^{{\rm(d)}\mu\nu}(p)
\right\vert_{\rm div}
=
g^{\mu\nu}\,\frac{N_f}{2(4\pi)^2}
\left[
  4 \Lambda^2
  - 8 \bar{M}_V^2\, \ln \Lambda^2
\right]
\ ,
\nonumber\\
&&
\left.
\Pi_{\overline{V}\overline{V}}^{{\rm(e)}\mu\nu}(p)
\right\vert_{\rm div}
=
g^{\mu\nu}\,\frac{N_f}{2(4\pi)^2}
\left[
  - 4 \Lambda^2
  + 4 \bar{M}_V^2\, \ln \Lambda^2
\right]
\ ,
\nonumber\\
&&
\left.
\Pi_{\overline{V}\overline{V}}^{{\rm(f)}\mu\nu}(p)
\right\vert_{\rm div}
=
g^{\mu\nu}\,\frac{N_f}{2(4\pi)^2}
\left[
  - \frac{1}{2} \Lambda^2
  + \frac{1}{2} \bar{M}_V^2\, \ln \Lambda^2
\right]
\nonumber\\
&&\quad
{}- \left( g^{\mu\nu} p^2 - p^\mu p^\nu \right)
\, \frac{N_f}{2(4\pi)^2} \frac{1}{12} \ln \Lambda^2
\nonumber\\
&&
\left.
\Pi_{\overline{V}\overline{V}}^{{\rm(g)}\mu\nu}(p)
\right\vert_{\rm div}
=
g^{\mu\nu}\,\frac{N_f}{2(4\pi)^2}
\left[
  - \frac{a^2}{2} \Lambda^2
\right]
\nonumber\\
&&\quad
{}- \left( g^{\mu\nu} p^2 - p^\mu p^\nu \right)
\, \frac{N_f}{2(4\pi)^2} \frac{a^2}{12} \ln \Lambda^2
\ .
\label{VV2p}
\end{eqnarray}
\noindent
Summing up the contributions in Eq.~(\ref{VV2p}), we obtain the
following divergent contribution:
\begin{eqnarray}
&&
\left.
\Pi_{\overline{V}\overline{V}}^{%
{\mbox{\scriptsize(1-loop)}}\mu\nu}(p)
\right\vert_{\rm div}
\nonumber\\
&&\quad
=
- \frac{N_f}{4(4\pi)^2} \left[
  (1+a^2) \Lambda^2 + 3 a g^2 F_\pi^2 \ln \Lambda^2
\right]
g^{\mu\nu}
\nonumber\\
&& \qquad
{} + 
\frac{N_f}{2(4\pi)^2} \, \frac{87-a^2}{12} \ln \Lambda^2
\left( p^2 g^{\mu\nu} - p^\mu p^\nu \right)
\ .
\label{VV:div}
\end{eqnarray}
On the other hand, 
the tree contribution is given by
\begin{equation}
\Pi_{\overline{V}\overline{V}}^{{\rm(tree)}\mu\nu}(p^2)
= F_{\sigma,{\rm bare}}^2 \, g^{\mu\nu} 
- \frac{1}{g^2_{\rm bare}}
\left( p^2 g^{\mu\nu} - p^\mu p^\nu \right) \ .
\end{equation}
The first term in Eq.~(\ref{VV:div}) which is proportional to
$g^{\mu\nu}$ is renormalized by $F_{\sigma,{\rm bare}}^2$ through
the requirement in Eq.~(\ref{Fs:z1:renorm}).
The second term in Eq.~(\ref{VV:div}) is renormalized by 
$g_{\rm bare}$ through
\begin{equation}
\frac{1}{g^2_{\rm bare}} - 
\frac{N_f}{2(4\pi)^2} \, \frac{87-a^2}{12} \ln \Lambda^2
= \mbox{(finite)}
\ .
\end{equation}
This renormalization leads to the following RGE
for $g$ [the third equation in Eqs.~(\ref{RGE for g2})]:
\begin{equation}
\mu \frac{d g^2}{d\mu} = - 
\frac{N_f}{2(4\pi)^2}
\frac{87 - a^2}{6} g^4 \ .
\label{app:RGE:g}
\end{equation}

\begin{figure}[htbp]
\begin{center}
\epsfxsize = 8cm
\  \epsfbox{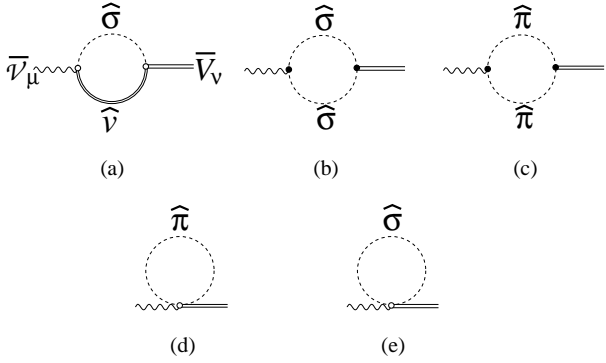}
\end{center}
\caption[$\overline{\cal V}_\mu$-$\overline{V}_\nu$ Two-Point 
function]{%
One-loop corrections to the two-point function
$\overline{\cal V}_\mu$-$\overline{V}_\nu$.
The vertices in (a) are from $\Sigma^{(\sigma V_\beta)}_{ab}$ and
$\Sigma^{(V_\alpha \sigma)}_{ab}$ in Eqs.~(\ref{Ysv}) and (\ref{Yvs});
the vertices in (b) are from $\Gamma^{(\sigma\sigma)}_{\mu,ab}$ in 
Eq.~(\ref{Xss}) together with derivatives acting on the quantum
fields;
the vertices in (c) are from $\Gamma^{(\pi\pi)}_{\mu,ab}$ in 
Eq.~(\ref{Xpp}) together with derivatives acting on the quantum
fields;
the vertex in (d) is from the second term of $\Sigma^{(\pi\pi)}_{ab}$
in Eq.~(\ref{Ypp}) and $\sum_c \Gamma^{(\pi\pi)}_{\mu,ac} 
\Gamma^{\mu,(\pi\pi)}_{cb}$;
the vertex in (e) is from the second term of 
$\Sigma^{(\sigma\sigma)}_{ab}$ in 
Eq.~(\ref{Yss}) and $\sum_c \Gamma^{(\sigma\sigma)}_{\mu,ac} 
\Gamma^{\mu,(\sigma\sigma)}_{cb}$.
}\label{fig:vr}
\end{figure}
We also calculate the one-loop correction to the two-point
function $\overline{\cal V}_\mu$-$\overline{V}_\nu$ to determine the
renormalization of $z_3$.
The relevant diagrams are shown in Fig.~\ref{fig:vr}.
The divergent contributions are evaluated as
\begin{eqnarray}
&&
\left.
\Pi_{\overline{\cal V}\overline{V}}^{{\rm(a)}\mu\nu}(p)
\right\vert_{\rm div}
=
g^{\mu\nu}\,\frac{N_f}{2(4\pi)^2}
\left[
  2 \bar{M}_V^2\, \ln \Lambda^2
\right]
\ ,
\nonumber\\
&&
\left.
\Pi_{\overline{\cal V}\overline{V}}^{{\rm(b)}\mu\nu}(p)
\right\vert_{\rm div}
=
g^{\mu\nu}\,\frac{N_f}{2(4\pi)^2}
\left[
  - \frac{1}{2} \Lambda^2
  + \frac{1}{2} \bar{M}_V^2\, \ln \Lambda^2
\right]
\nonumber\\
&&\quad
{}- \left( g^{\mu\nu} p^2 - p^\mu p^\nu \right)
\, \frac{N_f}{2(4\pi)^2} \frac{1}{12} \ln \Lambda^2
\ ,
\nonumber\\
&&
\left.
\Pi_{\overline{\cal V}\overline{V}}^{{\rm(c)}\mu\nu}(p)
\right\vert_{\rm div}
=
- g^{\mu\nu}\,\frac{N_f}{2(4\pi)^2} \, \frac{a(2-a)}{2} \Lambda^2
\nonumber\\
&&\quad
{}- \left( g^{\mu\nu} p^2 - p^\mu p^\nu \right)
\, \frac{N_f}{2(4\pi)^2} \frac{a(2-a)}{12} \ln \Lambda^2
\ ,
\nonumber\\
&&
\left.
\Pi_{\overline{\cal V}\overline{V}}^{{\rm(d)}\mu\nu}(p)
\right\vert_{\rm div}
=
g^{\mu\nu}\,\frac{N_f}{2(4\pi)^2} a \Lambda^2
\ ,
\nonumber\\
&&
\left.
\Pi_{\overline{\cal V}\overline{V}}^{{\rm(e)}\mu\nu}(p)
\right\vert_{\rm div}
=
g^{\mu\nu}\,\frac{N_f}{2(4\pi)^2}
\left[
  \Lambda^2
  - 2 \bar{M}_V^2\, \ln \Lambda^2
\right]
\ .
\end{eqnarray}
Thus
\begin{eqnarray}
&&
\left.
\Pi_{\overline{\cal V}\overline{V}}^{%
{\mbox{\scriptsize(1-loop)}}\mu\nu}(p)
\right\vert_{\rm div}
\nonumber\\
&&\quad
=
\frac{N_f}{4(4\pi)^2} \left[
  (1+a^2) \Lambda^2 + 3 a g^2 F_\pi^2 \ln \Lambda^2
\right]
g^{\mu\nu}
\nonumber\\
&& \qquad
{} -
\frac{N_f}{2(4\pi)^2} \, \frac{1+2a-a^2}{12} \ln \Lambda^2
\left( p^2 g^{\mu\nu} - p^\mu p^\nu \right)
\ .
\label{Vr:div}
\end{eqnarray}
The tree contribution is given by
\begin{equation}
\Pi_{\overline{\cal V}\overline{V}}^{{\rm(tree)}\mu\nu}(p^2)
= F_{\sigma,{\rm bare}}^2 \, g^{\mu\nu} 
+ 2 z_{3,{\rm bare}}
\left( p^2 g^{\mu\nu} - p^\mu p^\nu \right) \ .
\end{equation}
The first term in Eq.~(\ref{Vr:div}) which is proportional to
$g^{\mu\nu}$ is renormalized by $F_{\sigma,{\rm bare}}^2$ through the
requirement in Eq.~(\ref{Fs:z1:renorm}).
The second term in Eq.~(\ref{Vr:div}) is renormalized by 
$z_{3,{\rm bare}}$ through
\begin{equation}
z_{3,{\rm bare}} 
- 
\frac{N_f}{2(4\pi)^2} \, \frac{1+2a-a^2}{12} \ln \Lambda^2
= \mbox{(finite)}
\ .
\end{equation}
This leads to [the third equation in Eqs.~(\ref{RGE z})]
\begin{equation}
\mu \frac{d z_3}{d\mu} =
\frac{N_f}{(4\pi)^2} \frac{1+2a-a^2}{12} \ .
\label{app:RGE:z3}
\end{equation}

To summarize, Eqs.~(\ref{app:RGE:Fp}), (\ref{app:RGE:a}) and
(\ref{app:RGE:g}) are the RGE's for
$F_\pi^2$, $a$ and $g$ shown in Eq.~(\ref{RGE for g2}), and
Eqs.~(\ref{app:RGE:z1}), (\ref{app:RGE:z2}) and (\ref{app:RGE:z3})
are the RGE's for $z_1$, $z_2$ and $z_3$
shown in Eq.~(\ref{RGE z}).

Below the $m_\rho$ scale, 
$\rho$ decouples 
and hence $F_\pi$ runs by the loop effect of $\pi$ alone.
The relevant Lagrangian with least derivatives is given by
the first term of Eq.~(\ref{leading ChPT}) 
[or equivalently, the first term of Eq.~(\ref{Lagrangian})],
and the diagram contributing to $F_\pi^2$ is shown in
Fig.~\ref{fig:aa}(c).
The resultant RGE for $F_\pi$ is given by
\begin{equation}
\mu \frac{ d }{d\mu} \left[F_\pi^{(\pi)}\right]^2
= \frac{2N_f}{(4\pi)^2} \mu^2 
\qquad (\mu < m_\rho) \ .
\end{equation}
Unlike the parameters renormalized in a mass independent scheme, the
parameter $F_\pi(\mu)$ ($\mu<m_\rho$) does not smoothly
connect to $F_\pi(\mu)$ ($\mu > m_\rho$) at the $m_\rho$ scale.
We need to include the effect of finite renormalization.
This is evaluated by taking quadratic divergence proportional to $a$
in Eq.~(\ref{div:aa}) and replacing $\Lambda$ by $m_\rho$.
This leads to the relation (\ref{finite renormalization}):
\begin{equation}
\left[ F_\pi^{(\pi)}(m_\rho) \right]^2 =
F_\pi^2(m_\rho) + 
\frac{N_f}{(4\pi)^2} \frac{a(m_\rho)}{2} m_\rho^2 \ ,
\end{equation}
where $F_\pi^{(\pi)}(\mu)$ runs by the loop effect of $\pi$ alone for
$\mu<m_\rho$.

Finally, let us show the finite correction to the relation for
$L_{10}$ given in Eq.~(\ref{l10}).
This is evaluated from the finite part of the $g^{\mu\nu}$ part of the
$\overline{\cal A}_\mu$-$\overline{\cal A}_\nu$ two-point function.
[Here the $g^{\mu\nu}$ part of the
$\overline{\cal A}_\mu$-$\overline{\cal A}_\nu$ two-point function
is defined by $\Pi_{\overline{\cal A}\overline{\cal A}}^{L}(p^2)
\equiv
\frac{p_\mu p_\nu}{p^2} 
\Pi_{\overline{\cal A}\overline{\cal A}}^{\mu\nu}(p)$.]
{}From Fig.~\ref{fig:aa} we obtain
\begin{eqnarray}
&&
\Pi_{\overline{\cal A}\overline{\cal A}}^{(a)L}(p)
=
- N_f a \bar{M}_v^2 B_0(p^2;\bar{M}_v,0)
\ ,
\nonumber\\
&&
\Pi_{\overline{\cal A}\overline{\cal A}}^{(b)L}(p)
=
N_f \frac{a}{4} \left[
  B_A(p^2;\bar{M}_v,0) - A_0(\bar{M}_v) - A_0(0)
\right]
\ ,
\nonumber\\
&&
\Pi_{\overline{\cal A}\overline{\cal A}}^{(c)L}(p)
=
N_f (a-1) A_0(0)
\ ,
\end{eqnarray}
where
\begin{eqnarray}
&&
A_0(M^2) \equiv
\int \frac{d^nk}{i(2\pi)^n}
\frac{1}{M^2-k^2}
\ ,
\nonumber\\
&&
B_0(p^2;M,m) \equiv
\int \frac{d^nk}{i(2\pi)^n}
\frac{1}{ [M^2-k^2] [m^2-(k-p)^2] }
\ ,
\nonumber\\
&&
B_A(p^2;M,m) \equiv
\nonumber\\
&&\quad
\frac{(M^2-m^2)^2}{p^2}
\left[ B_0(p^2;M,m) - B_0(0;M,m) \right]
\ .
\end{eqnarray}
According to the analysis in Ref.~\cite{Tanabashi}, the
${\cal O}(p^2)$ part of
$\Pi_{\overline{\cal A}\overline{\cal A}}%
^{{\mbox{\scriptsize(1-loop)}}L}(p^2) \equiv
\Pi_{\overline{\cal A}\overline{\cal A}}^{(a)L}(p^2) +
\Pi_{\overline{\cal A}\overline{\cal A}}^{(b)L}(p^2) +
\Pi_{\overline{\cal A}\overline{\cal A}}^{(c)L}(p^2)$
gives a finite order correction to $L_{10}$ as
\begin{equation}
-\frac{1}{4}
\left.
\frac{d}{dp^2}
\Pi_{\overline{\cal A}\overline{\cal A}}%
^{{\mbox{\scriptsize(1-loop)}}L}(p^2)
\right\vert_{p^2=0}
= \frac{N_f}{(4\pi)^2} \frac{11a}{96}
\ ,
\label{l10fin}
\end{equation}
which is the last term in Eq.~(\ref{l10}).


\begin{thebibliography}{99}
\bibitem{foot:mudep}
Such dependences are assigned to the parameters $z_2(\Lambda)$ and
$z_1(\Lambda)$.
This situation is similar to that for the high energy parameters in
chiral perturbation theory~\cite{GL}.
\bibitem{foot:OPE}
It should be noticed that the
$\alpha_s/\pi$ term and $\alpha_s \left\langle \bar{q} q
\right\rangle^2$ term depend on the renormalization point $\mu$ of
QCD, 
and that those generate a small dependence of the bare parameters of
the HLS on $\mu$.
It is reasonable to take $\mu$ to be equal to the matching scale
$\Lambda$.
\bibitem{foot:piloop}
The RGE for $F_\pi^2$ for $\mu<m_\rho$ is given by putting $a=0$ in
the first equation of Eqs.~(\ref{RGE for g2}).
\bibitem{foot:mix}
Note that the existence of
kinetic type $\rho$-$\gamma$ mixing from the $z_3$ term was needed to
explain the experimental data of 
$\Gamma(\rho\rightarrow e^+e^-)$~\cite{HSc}.
\bibitem{foot:Borel}
One might think of the matching by the Borel
transformation of the correlators.
However, agreement of the predicted values, especially $g_\rho$, 
is not as remarkably good as that for the present case.
\bibitem{foot:rpp}
The contribution from the higher derivative term is neglected
in the expression of $g_{\rho\pi\pi}$ given in 
Eq.~(\ref{g rho pi pi}), i.e.,
$g_{\rho\pi\pi} = g_{\rho\pi\pi}(m_\rho^2;0,0) =
g_{\rho\pi\pi}(0;0,0)$.
\bibitem{foot:log}
The parameter choice $\Lambda=m_\rho$ does not work, either.
\bibitem{foot:Yss}
One might think that there exists a $\sigma$-tadpole contribution to
the $\overline{\cal A}$-$\overline{\cal A}$ two-point function with
the vertex from the 
second term of $\Sigma^{(\sigma\sigma)}_{ab}$ in Eq.~(\ref{Yss}).
However, the vertex is exactly canceled with the vertex from
$\sum_c \Gamma^{(\sigma\pi)}_{\mu,ac} 
\Gamma^{\mu,(\pi\sigma)}_{cb}$.
Thus there is no $\sigma$-tadpole contribution to
the $\overline{\cal A}$-$\overline{\cal A}$ two-point function.
Similar cancellations occur
between contributions from the first term
of $\Sigma^{(\sigma\sigma)}_{ab}$
and $\sum_c \Gamma^{(\sigma\sigma)}_{\mu,ac} 
\Gamma^{\mu,(\sigma\sigma)}_{cb}$ to 
$\overline{\cal V}$-$\overline{\cal V}$
as well as to 
$\overline{V}$-$\overline{V}$.
\end{thebibliography}
\end{document}